\documentclass{article}
\usepackage[utf8]{inputenc} 
\usepackage{amsmath, amssymb} 
\usepackage{graphicx}
\usepackage{svg}
\usepackage[margin=1in]{geometry}
\usepackage{multirow}
\usepackage{array}
\usepackage{caption}
\usepackage{tabularx}
\usepackage{wrapfig}   
\usepackage{subcaption}
\usepackage[normalem]{ulem}
\usepackage{tikz}

\title{
Physical scaling laws in dislocation microstructures and avalanches from dislocation dynamics simulations
}
\author{M. Aissaoui, C. Kahloun, O.U. Salman, S. Queyreau}

\begin{document}

\maketitle

\begin{abstract}
{Avalanche-like plastic bursts in crystalline materials follow power law statistics, but the scaling exponents and cutoff parameters vary widely in the literature ($\alpha$ ranging from 1 to 2.2), hindering predictive modeling.  Since distributions do not follow Gaussian behavior, the average of plastic kinetics is not correctly defined. Larger-scale models that rely on average behavior are therefore fundamentally flawed. {We performed extensive three-dimensional Dislocation Dynamics simulations} of FCC Cu deformation across three orders of magnitude in dislocation density ($\rho = 5 \times 10^{10} \ \text{to} \ 2 \times  10^{12} \ \text{m}^{-2}$) under constant strain rates. Our results demonstrate that the power law exponent ($\alpha \approx  1.6 \pm 0.1$ ) is invariant to both dislocation density and loading direction, resolving previous inconsistencies. However, dislocation density strongly controls the power law truncation scaling ($\Delta \gamma_{max} \propto \ b/\sqrt{\rho}$) and the distribution of avalanche triggering stresses. We quantify correlations between slip system activities and show how individual system contributions evolve with avalanche size. These findings reconcile experimental scatter in avalanche statistics and provide quantitative scaling laws for mesoscale-to-continuum plasticity models.}
\end{abstract}

\section*{Introduction}

{While plastic deformation appears smooth at engineering scales, it proceeds through discrete, avalanche-like bursts at the microscale.} Plastic events $\Delta \gamma$ follow the so-called power law distribution $P(\Delta \gamma)\propto \Delta \gamma^{-\alpha}$. These avalanche statistics can be obtained through  analysis of the discrete part of acoustic emission signals \cite{weiss1997acoustic, richeton2005breakdown, weiss2019plastic}, or from the strain increments seen on the deformation curves obtained from lower scale simulations \cite{zaiser2006scale, devincre2008dislocation, csikor2007dislocation} or deformed micropillars \cite{dimiduk_scale-free_2006, sparks2018nontrivial, alcala_statistics_2020}. 

While power law (PWL) exponents found for 3D dislocation-based plasticity of fcc materials can well match those obtained for other non-equilibrium phenomena, the range of reported values is rather large, from 1 to 2.2 \cite{ richeton2005breakdown, devincre2008dislocation, csikor2007dislocation, dimiduk_scale-free_2006, sparks2018nontrivial, weiss_mild_2015, devincre2010scale}. It is therefore unsure whether dislocation avalanches correspond to a single universality class \cite{sparks2018nontrivial, weiss_mild_2015}. {Power law exponent is shown to depend strongly upon the material and crystal structure \cite{sparks2018nontrivial}, the type of loading control \cite{csikor2007dislocation, cui_controlling_2016}, imposed strain rate \cite{sparks2018nontrivial, sparks_avalanche_2019}, and possibly single-crystal orientation \cite{sparks2018nontrivial}. Assessing the organization and strength of the dislocation microstructure through the triggering stresses at the origin of avalanches could help in understanding the variety of critical exponents seen, however this analysis is typically hard to achieve in practice \cite{zaiser2006scale, berta2025identifying}. {More systematic studies are thus required to rigorously define what controls the power law exponent for dislocation avalanches and to conclude on the universality class to which they belong.}}

{The pure power law distribution is known to be scale free, and a truncated power law} of the form $P(\Delta \gamma)\propto \Delta \gamma^{-\alpha} \exp(-\Delta \gamma/\Delta \gamma_{max})$ is often employed to model the decreasing contribution of the largest plastic events \cite{csikor2007dislocation, sparks_avalanche_2019}. {Similarly to the exponent, the power law truncation may be affect by many parameters} \cite{csikor2007dislocation} like the amount of deformation \cite{sparks2018nontrivial, weiss_mild_2015}, single-crystal orientation \cite{devincre2008dislocation, sparks2018nontrivial, devincre2010scale} and imposed strain rate \cite{csikor2007dislocation, cui_controlling_2016, sparks_avalanche_2019}.{ The determination of the power law cutoffs is crucial in defining the average behavior and thus rigorously linking the mesoscopic picture of dislocation mechanisms to the continuous behaviour of materials at the macroscale. Without this, the average plastic behaviour is ill-defined preventing any quantitative upscaling. The truncation behavior of the power law requires significant data, perhaps even more so than the identification of the power law exponent, in particular for the rare events corresponding to large avalanches.} Power law truncation can make exponent identification more difficult, and its description is not yet fully quantitative.

{Loading orientation may strongly impact the type of dislocation interactions, their strength, and resulting organization in dislocation microstructures. Changing the relative strengths and internal lengths of strengthening mechanisms has been shown to dramatically impact the statistical signature of plastic bursts in Al alloys micropillars} \cite{Zhang2017}. However the impact of orientation has been rarely studied  \cite{sparks2018nontrivial, devincre2010scale}, and different consequences were observed. 
{In \cite{devincre2010scale}, a power-law exponent of $\alpha \approx 1.7$ was found to be independent of loading orientation in Discrete Dislocation Dynamics (DDD) simulations conducted at a fixed strain rate, whereas experiments on Au single-crystal micropillars with different orientations reported $\alpha \approx 1.6-1.9$ for high-symmetry axes and $\alpha \approx 1-1.2$ for low-symmetry axes \cite{sparks2018nontrivial}.}
{Ultimately, one may wonder what controls the dislocation avalanche statistics in deformed systems with evolving microstructures, and polycrystalline samples with {multiple} crystal orientation.}

Understandably, many existing experimental and simulation works focus on deformed micropillars in which the intermittency of plastic deformation is exacerbated. However, conventional dislocation mechanisms compete with the size effects induced in these finite-dimension systems \cite{el-awady_unravelling_2015}. On a macroscopic scale, power law regimes and corresponding exponents are well defined over amplitudes or energies of up to five orders of magnitude from the analysis of acoustic emission bursts \cite{weiss1997acoustic, richeton2005breakdown, weiss_mild_2015}, although the link with dislocation activity is not direct. DDD simulations appear to be the simulation tool of choice, but few studies have focused on bulk systems in 3D \cite{devincre2008dislocation, csikor2007dislocation, devincre2010scale, berta2025identifying}. 

{In this work, we show the changes in the dislocation avalanche statistics simulated by 3D Dislocation Dynamics simulations in the pure Cu bulk system, which represents the fundamental underpinning of more complex systems.} We focus on two \emph{microstructural} parameters, namely: the dislocation density and dislocation microstructure through the crystal orientation, that may change material structure (disorder) and relative contributions of short- long-range contributions. Thanks to a careful simulation design, a large amount of data is obtained, enabling PWL regimes of up to four orders of magnitude to be unambiguously defined, bounded by well-defined cutoffs, clarifying some of the inconsistencies in the literature. Avalanche data are analyzed in greater detail to provide a unique picture of plastic deformation, such as the contribution of different slip systems, the distribution of critical configurations at the onset of avalanches and the correlation between systems. Characterization of the statistics enables quantitative modeling of avalanche statistics at the mesoscale, paving the way for new comparisons with experiments.

\section{Methodology}

We conducted a comprehensive investigation using large scale DDD simulations with the $microMegas$ code, which has been described in details in 
\cite{devincre2011modeling,queyreau2020dislocation} and making use of the Cai et al.'s non-singular elastic theory \cite{cai2006non,arsenlis2007enabling,queyreau2014analytical}. This section only covers the specific conditions for the simulations carried out here. 

{ All simulations are performed in 3D to be able to capture junction formation, cross-slip activity and microstructure formation. The figure \ref{fig:supp-avalan} illustrates an avalanche sequences and the figure \ref{fig:extendedplane} provides a sense of the simulation domain size.} In this study, we performed a series of model simulations of the tensile deformation of Cu single-crystals oriented along different directions leading to different slip activities and typical microstructures.  Simulations encompass from [135] single-slip activation leading to dislocation entanglements, [112] double slip condition leading to planar microstructures, a latent hardening experiment \cite{devincre2006physical,queyreau2009slip} for which the forest density is immobile and [001] stable multislip conditions building cells. This latter orientation is the focus of our study, as multislip condition is the dominant slip condition existing in polycrytals deformed at large strain. The elastic behavior of Cu is assumed to be linear and isotropic with a shear modulus $\mu = 42$ GPa and Poisson’s ratio $\nu = 0.34$. 

In the absence of in-depth knowledge of the  Frank's initial network, the initial microstructure consists of a random distribution of dislocation prismatic loops with a length of $ 4\ \mu m$ for segments in the primary systems and $ 1\ \mu m$  for segments in the corresponding collinear system (also referred to cross-slip systems). {The choice of simulation box size $L_B$ represents a trade-off between (i) physical relevance—since a larger box can accommodate a greater diversity of dislocation configurations—and (ii) numerical efficiency, as computational cost increases with the number of dislocation segments. In this work, the box size was set to $L_B =10 \Lambda$, following common practice in recent DDD simulations, where $\Lambda \approx 1/ \sqrt{\rho_0}$ represents the mean dislocation spacing for a density $\rho_0$, in accordance with the similitude principle that governs most dislocation interactions and microstructural features. Depending on the initial density, the box sizes thus range from $(7 \mu m)^3$ to $(44 \mu m)^3$. }

{For [001] deformation, plastic deformation experimentally occurs on four (out of eight possible) systems. In the corresponding simulations, following \cite{devincre2007collinear}, initial density is set larger on four \emph{primary} systems (collinear to the other four remaining systems) so as to obtain the typical hardening with stable deformation along the [001] axis. We checked that the avalanche statistics is unchanged when starting with identical densities on all possible systems.} Several initial realizations of the microstructures are considered, our analysis is limited to the simulations deemed the most realistic (e.g. with well-balanced slip activity leading to a stable loading direction). Periodic boundary conditions (PBC) are applied. Cross-slip for screw dislocations is controlled by a KMC algorithm following the model in \cite{kubin2013dislocations,kubin2009dynamic}. {Theses arbitrary initial microstructures quickly evolve into realistic dislocation arrangements (see Fig. \ref{fig:supp-avalan})} and this design has been successfully employed to quantify dislocation storage \cite{devincre2008dislocation} or plastic hysteresis \cite{queyreau2021multiscale}. 

The dislocation density is varied over three orders of magnitude, starting from $\rho_0 = 5\times 10^{10}$ to $2\times10^{12} \ m^{-2}$ and deformation during a simulation typically leads to a significant increase of density of up to an order of magnitude, further widening the range of dislocation density considered here. Dislocation density is changed in two ways: i) by starting with different microstructures while preserving the box dimensions, ii) by preserving the initial structure but resizing the entire geometry (and keeping $L_B = 10 \Lambda$). Here, deformation is driven using a constant plastic strain rate $\dot{\epsilon}_a$, similar to strain driven experiments, with an infinite rigid machine stiffness. We carefully chose $\dot{\epsilon}_a$ = 50 s$^{-1}$ so as to preserve numerical efficiency while avoiding strain rate effects. In other words, deformation is solely controlled by forest interactions \cite{fan2021strain}. Our simulations reveal that the imposed strain rate has a dramatic impact on the avalanche statistics also seen in pillars \cite{sparks2018nontrivial, sparks_avalanche_2019}. {We found that the loading rate $\dot{\epsilon}_a$ has a rather complex and subtle impact on the  plastic deformation at the mesoscale and dislocation avalanches' signature, and this will be the subject of a forthcoming paper } \cite{aissaoui2025connecting}

Great care has been paid to extract the most of and precise data from the simulations as significant statistics are required to correctly define avalanche statistics. {To achieve this, we opted for two changes compared to what is done in other simulation studies. i) Microstructure properties were saved at high frequency and high floating point precision. ii) The use of PBC allows for simulating larger deformation with many more plastic events.} However, the presence of PBC may impact the largest plastic events. To correctly capture the natural maximal extension of avalanches and power law truncation of avalanche statistics, we optimized the ratios of simulation box dimensions so as to obtain a very large extended slip plane -corresponding to the various tiles of slip plane swept by a dislocation loops while expanding (see supplementary material). {Under single-slip conditions, plastic activity is dominated by the primary slip system, while the colinear cross-slip system contributes little due to its lower Schmid factor. In the multislip case, primary and colinear slip systems share identical Schmid factors, allowing all eight systems to accommodate plastic deformation. Consequently, the swept area per slip system required to achieve the same total strain is smaller in multislip than in single-slip conditions.} For an individual slip system, the largest sweapt area free of any PBC artifacts, can be as large as $\pi(65 \mu m)^2$ for our multislip simulations and $\pi(95 \mu m)^2$ for the single slip simulations. We will see that our simulation design allows large data sets, with about 20,000 plastic events per simulation, and well-defined statistical signatures.  {Simulations are terminated when the computation time for a single time step exceeds one minute. The total computational cost of a single simulation is approximately 40,000 CPU·h. More details regarding simulations are provided in supplementary material.}

Even with sufficient data, analyzing avalanche statistics can be a difficult task. Here, we reemploy and built upon the careful methodology derived by Clauset and coworkers \cite{clauset2009power}. To provide a quantitative description of the avalanche statistics from the simulations, most of our data is fitted using the robust maximum likelyhood methodology by a truncated power law with an exponential decay (see supplementary material). The fits are based on data truncated beyond the lower $\Delta \gamma_{min}$ limit for the PWL regime.  Curves (dashed lines) and fit parameters are generally shown in the figures.

\section{Results}

\subsection{Impact of the dislocation density on avalanche statistics}

\begin{figure}[h]
    \centering
    \begin{subfigure}[b]{0.45\textwidth}
        \centering
        \includegraphics[width=\textwidth]{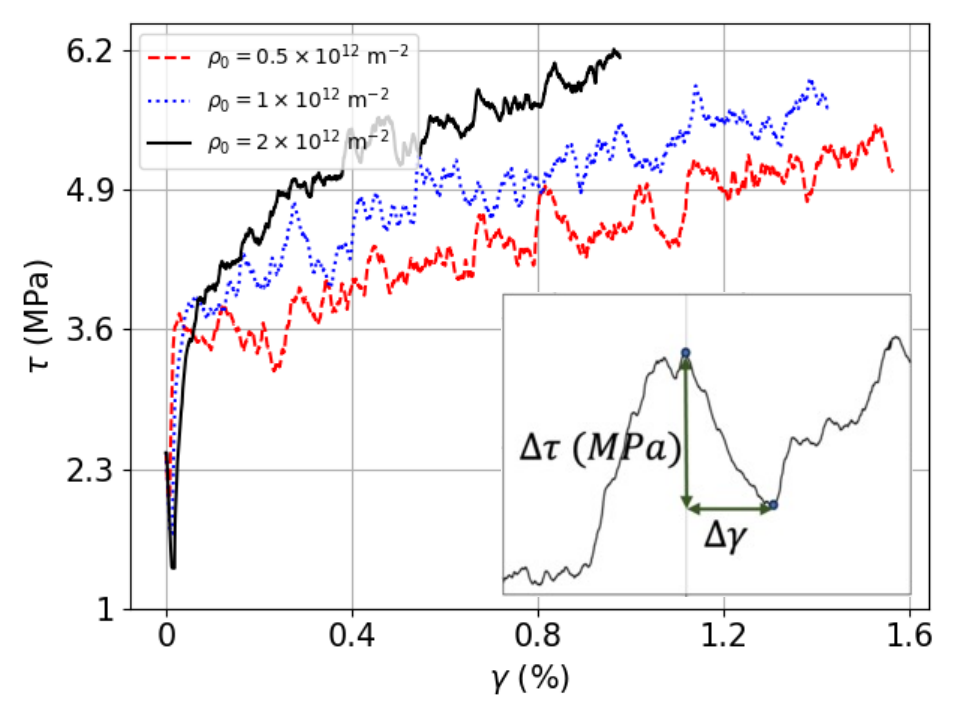}
        \caption{}
    \end{subfigure}
    \hfill
    \begin{subfigure}[b]{0.45\textwidth}
        \centering
        \includegraphics[width=\textwidth]{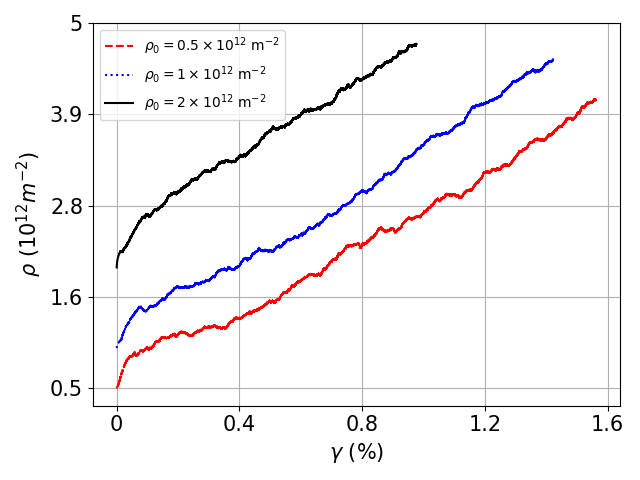}
        \caption{}
    \end{subfigure}
    \vfill
    \begin{subfigure}[b]{0.45\textwidth}
        \centering
        \includegraphics[width=\textwidth]{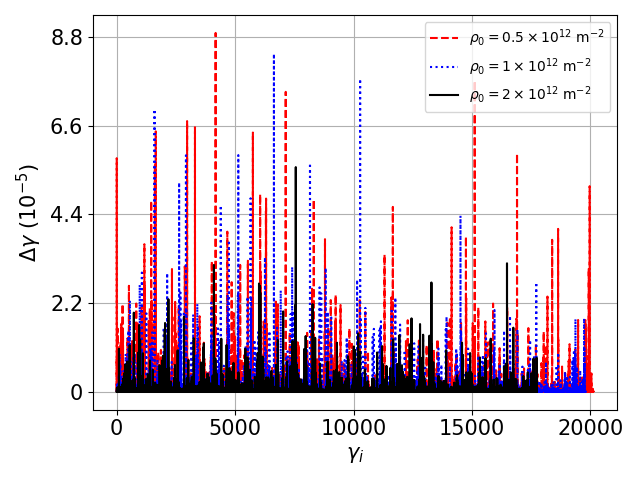}
        \caption{}
    \end{subfigure}
    \hfill
    \begin{subfigure}[b]{0.45\textwidth}
        \centering
        \includegraphics[width=\textwidth]{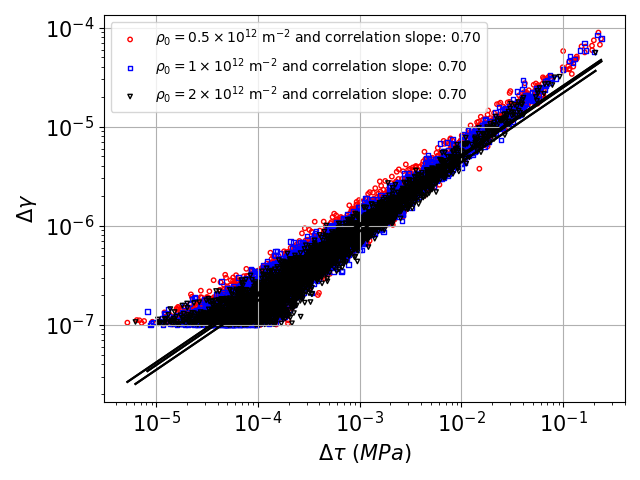}
        \caption{}
    \end{subfigure}
    \caption{Typical mesoscopic plastic behavior simulated by DDD for Cu [001] single crystals under constant strain rate. (a) Deformation curves with the shear stress \(\tau\) as function of the total shear strain \(\gamma\) for different initial density \(\rho_0\). Inset: zoom in on one of the serrations seen on the deformation curve. We define the strain-bursts \(\Delta\gamma\) and the stress-drop \(\Delta\tau\) for every plastic events. (b) Corresponding evolutions of the dislocation density \(\rho\) as function of the total shear strain \(\gamma\). (c) Amplitude of individual strain-bursts \(\Delta\gamma)\). (d) Correlation between the strain-bursts \(\Delta\gamma\) and the stress-drop \(\Delta\tau\), for \(\Delta\gamma > 1\times10^{-7}\), well into the power law regime. {The legend also shows the values for  the correlation exponent $1/\beta$ (see main text).}} 
    \label{fig:defcurve001}
\end{figure}

Figure \ref{fig:defcurve001}.a) shows some of the deformation curves obtained for the [001] orientation and different dislocation densities. The plastic behaviour is typical of pure Cu single crystals at room temperature \cite{ kubin2009dynamic, takeuchi1973temperature}, with low flow stresses, a constant hardening rate of about $\mu/150$, and \emph{normal} dislocation storage with $\sqrt{\rho} \propto \gamma$ the shear strain (cf. Fig. \ref{fig:defcurve001}.b)) with very little dynamic recovery. In what follows, avalanche-like plastic bursts are characterized by a stress drop $\Delta \tau$ and strain increment $\Delta \gamma$ (see inset of Fig. \ref{fig:defcurve001}.a)). Fig. \ref{fig:defcurve001}.c) shows the amplitude of discrete events $\Delta \gamma_i$, whose number increase during microyielding and reaches up to 20,000 events at the end of each single simulation. 

\begin{figure}[ht]
    \centering
    \begin{subfigure}[b]{0.45\textwidth}
        \centering
        \includegraphics[width=\textwidth]{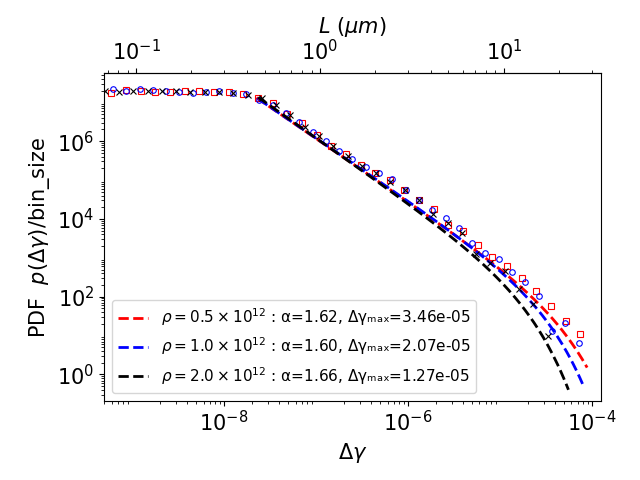}

    \end{subfigure}
    \hfill
    \begin{subfigure}[b]{0.45\textwidth}
        \centering
        \includegraphics[width=\textwidth]{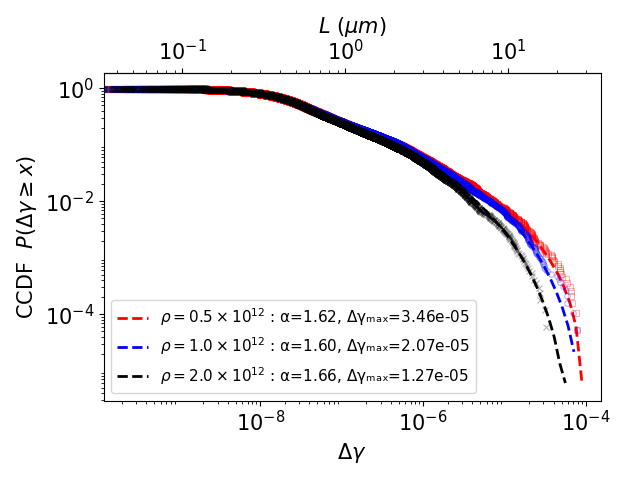}

    \end{subfigure}

    \centering
    \begin{subfigure}[b]{0.45\textwidth}
        \centering
        \includegraphics[width=\textwidth]{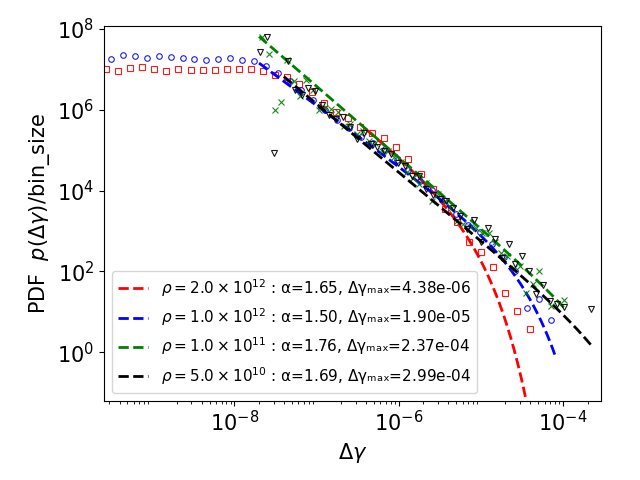}

    \end{subfigure}
    \hfill
    \begin{subfigure}[b]{0.45\textwidth}
        \centering
        \includegraphics[width=\textwidth]{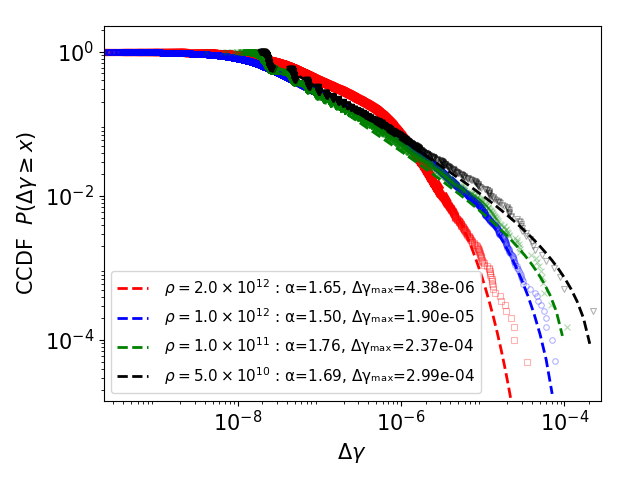}

    \end{subfigure}
    
    \caption{Impact of the dislocation density upon the strain resolved avalanche statistics during [001] deformation of Cu single crystals. {(a) and (c) Probability density function (pdf) of strain-bursts  $\Delta\gamma$ ($=x$) }. The avalanche extension $L$ is also given in units of $\mu$m to give a sense of the size of simulated plastic events. Complementary stress resolved data are provide in Supp. Mat. (b) and (d) Complementary cumulative probability function of strain-bursts $\Delta\gamma$. For a) and b), the box dimensions are fixed and dislocation density is changed. For c) and d), box dimensions are resized changing the dislocation density while preserving the numerical efficiency of DDD simulations at large densities. {The figures also show the values for the parameters $\alpha_i$ and $\lambda_i$ appearing in the truncated power law modeling strain burst statistics (see main text for more details).} }
    \label{fig:pdf001density}
\end{figure}

Next, we discuss the typical dislocation avalanche statistics focusing on plastic bursts $\Delta \gamma$ for reasons that will be explained later (stress resolved $\Delta \sigma$ analysis can be found in the Supp. Mat.). Figure \ref{fig:pdf001density}.a) and .c) show the probability density function (pdf) of $\Delta \gamma$ as function of the initial dislocation density. The statistics exhibits an unambiguous power law regime $p(\Delta \gamma) \propto \Delta \gamma^{-\alpha}$ over a domain of up to four orders of magnitude in event size, which is rarely seen in simulations. The exponent $\alpha$ in the power law regime is about negative 1.6-1.7, similarly to what has been observed in previous 3D DDD in bulk systems \cite{devincre2008dislocation,devincre2010scale}, and in the range of what is observed in experiments on metallic pillars under compression \cite{sparks2018nontrivial, sparks_avalanche_2019} or deformed single crystals \cite{weiss_mild_2015}.

Figure \ref{fig:defcurve001}.d) shows that the stress drop $\Delta \tau$ and strain increment $\Delta \gamma$ are correlated, e.g. a large $\Delta \tau$ typically corresponds to a large $\Delta \gamma$. However, for small events, dispersions are very large. The correlation roughly follows $\Delta \tau \propto \Delta \gamma^\beta$ with $\beta = 1.43$ here. Interestingly, the exponent of correlation $\beta$ is unaffected by dislocation density nor loading conditions (see below). This constant correlation justifies our focus on strain resolved $\Delta \gamma$ data. 

Now, let us focus on the bounds delimiting the PWL regime. The cutoffs are best seen on the complementary cumulative distribution function ({CCDF}) shown in Fig. \ref{fig:pdf001density}.b) and d). To give a sense of the size of plastic events simulated, we define an avalanche extension metric $L$ defined from Orowan's law as $L^2 = \Delta \gamma Vol /b$, with  $Vol$ the simulated volume. The size of avalanches $\Delta \gamma$ defined in this metric is also provided in the figure \ref{fig:pdf001density}. Following the idea of Kubin and collaborators, the average of $<L>$ is expected to be connected to the so-called Mean Free Path of dislocations \cite{devincre2008dislocation}, although the explicit relationship has yet to be found. 

{
The statistics obtained from our carefully designed simulations present well defined power law bounds, which is not always the case in the literature. Here, the smallest strain bursts $\min(\Delta \gamma)$ are independent from the dislocation density and of about $3\times10^{-8}$. This is equivalent to $L^2 \approx (0.05 \mu m)^2$, which corresponds certainly to very small displacements of individual dislocations (see next section). The truncation behaviour associated to largest events, is also well defined in our analysis, and the cutoff parameter $\Delta \gamma_{max}$ is strongly impacted by the dislocation density as $\max({\Delta \gamma})$ goes from $3 \times 10^{-4}$ to $2 \times 10^{-5}$ when the initial dislocation density increases from $\rho_0 = 5 \times 10^{10} \ m^{-2}$ to $2 \times 10^{12} \ m^{-2}$. In other words, the size of the largest observed events decreases by an order of magnitude with the density of \emph{obstacle} dislocations. The maximal size of the simulated avalanches reaches $L^2 \approx (30 \mu m)^2$ (in the fixed-dimension boxes), which is several times the box dimensions yet smaller than the extended plane dimensions. We will thus consider that the measured $\max({\Delta \gamma})$ accurately represents the \emph{maximal natural extension} of dislocation avalanches. }

{A change in the avalanche cutoff has also been observed in pillars prestrained by varying amounts of deformation \cite{sparks2018nontrivial}, corresponding to different dislocation densities (see Discussion section).}

Here, our results cover both of our simulation design strategies : i) fixed-box dimensions with increasing dislocation densities and ii) resized simulation boxes. The impact of dislocation density on the power law bounds is found to be coherent and similar in both sets of simulations (when accounting for the difference in total strain reached), therefore independent of the modeling strategy. { In 2D simulations, homothetic box geometries are sometimes employed in the litterature \cite{zaiser2006scale} to vary dislocation density}, while preserving the initial relative positions of segments. Deformation is expected to occur from similar successions of configurations whose activation stress scales with the box dimensions. However, in massive 3D simulations such as those carried out here, junction stability and cross-slip activity may be rather different  when physical dimensions are rescaled. In consequence, the succession of plastic events is different even when starting from the same -albeit rescaled- dislocation microstructure. {We will see shortly that the avalanche statistics are unaffected by this modeling choice for a given dislocation density.} Simulations with homothetic box geometry are however more numerically efficient and make it easier to obtain significant deformations and densities.

\subsection{Influence of the loading direction on avalanche statistics}

In this second set of simulations, the simulated microstructures are modified by deforming single crystals along different orientations. The initial dislocation density is taken as $\rho_0 = 10^{12}$ m$^{-2}$. Several realizations of the initial microstructure were tested, we have kept the one associated with a plastic deformation balanced between the different active slip systems, leading to the hardening rate expected from experiments. For the sake of brevity, the deformation curves and the evolution of the dislocation density curves are collected in the supplementary materials.

\begin{figure}[ht]
    \centering
    \begin{subfigure}[b]{0.45\textwidth}
        \centering
        \includegraphics[width=\textwidth]{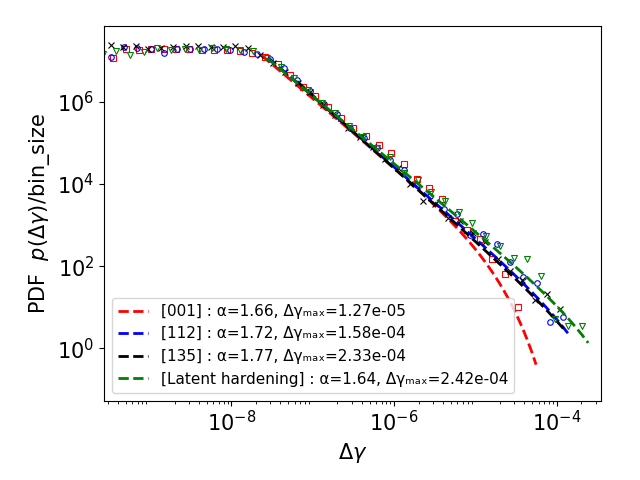}
        \caption{}
         
    \end{subfigure}
    \hfill
    \begin{subfigure}[b]{0.45\textwidth}
        \centering
        \includegraphics[width=\textwidth]{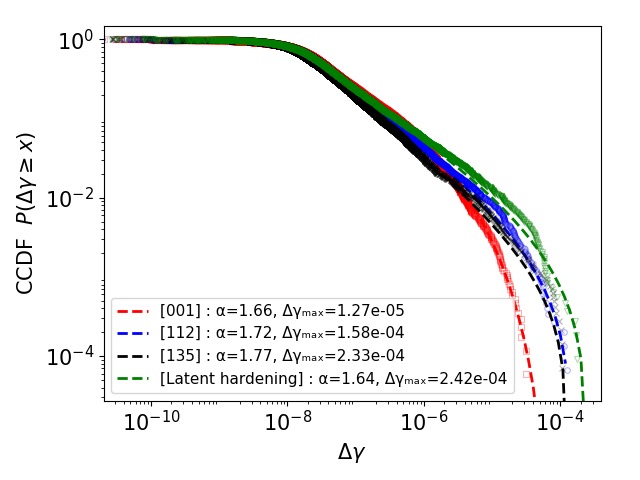}
        \caption{}
         
    \end{subfigure}
    \caption{Impact of the loading conditions on the avalanche statistics. (a) Probability density function of strain-bursts  \(\Delta\gamma\). (b) {CCDF} of strain-bursts \(\Delta\gamma \). The largest events may differ by about an order of magnitude depending on the loading condition, diminishing by as much the extend of the power law domain, while the power law exponent is mostly unchanged. 
    }
    \label{fig:pdforientation}
\end{figure}

Avalanche signatures displayed in Figure \ref{fig:pdforientation} for these new conditions share many of the features discussed previously. Irrespective of the orientation, an extended power law regime is found and associated to a similar exponent $\alpha \approx 1.7 \pm 0.1$. We focus on strain-resolved data because the approximate correlation between stress drop-strain burst is still operative $\Delta \tau \propto \Delta \gamma^\beta$ and the correlation exponent $\beta$ unchanged by loading direction. Stress resolved data can be found in the supplementary materials. Power law domain is delimited by well defined bounds.  The lower cutoff $\Delta \gamma_{min}$ seems to be slightly dependent upon the loading direction but more significant is the strong dependence of the upper truncation of the power law regime $\Delta \gamma_{max}$, which is clearly orientation dependent. $\Delta \gamma_{max}$ strongly decreases as the number of active slip system increases. In agreement with results in previous section, largest avalanches probability decreases as obstacle density increases.

A similar qualitative trend has been observed in a few of other studies from the literature. For example, Devincre and collab. found identical exponents for [135], [112], [111] and [001] deformed single crystals by means of 3D DDD simulations on bulk systems \cite{devincre2010scale}, the power law regime and its bounds are however less well-defined than here. { In their experimental study on compressed Au micropillars \cite{sparks2018nontrivial}, Sparks and Maaß reported orientation-dependent power-law exponents rather than a single constant value. The results can be grouped by crystallographic symmetry: for high-symmetry axes, $\alpha_{[001]} \approx 1.6$ (very similar to our findings), $\alpha_{[111]}\approx 1.9$, for axes with lower-symmetry $\alpha_{[123]} \approx 1$ and $\alpha_{[011]} \approx 1.2$}.
The difference with our findings is not yet fully understood, {but could be explained by i) the impact of finite size or surface effects in pillars of micrometric dimensions or ii) by the evolution of the cutoffs with dislocation density (see below)}. Next, we will present an analysis of what constitutes an avalanche in our DDD simulations so as to be able to quantify the evolution of the power law cutoff $\Delta \gamma_{max}$.

A decrease of the contribution of the largest avalanches as the density of obstacles blocking the dislocations increases is rather natural. However, this has two important consequences. i) In both simulations and experiments, dislocation density increases monotonically with accumulated deformation, which means that the resulting avalanche statistics are in fact associated to a constantly changing power law cutoff. ii) Given that avalanches occur on localized slip planes and different slip systems, one may wonder which part of the dislocation microstructure actually contributes to stopping avalanches.

\subsection{Contribution of individual slip systems and correlations in avalanches}

Here, we propose a complementary look at the statistical signature of avalanches on a per-system basis. Due to the shared statistical signatures of the phenomena belonging to the same class of universality, details of the physical mechanisms associated with avalanches are generally considered unimportant in the literature. However, these details of the underlying interactions may well become essential for characterizing power law truncations and provide valuable information for future larger-scale modeling.

\begin{figure}[h]
    \centering
        \includegraphics[width=\textwidth]{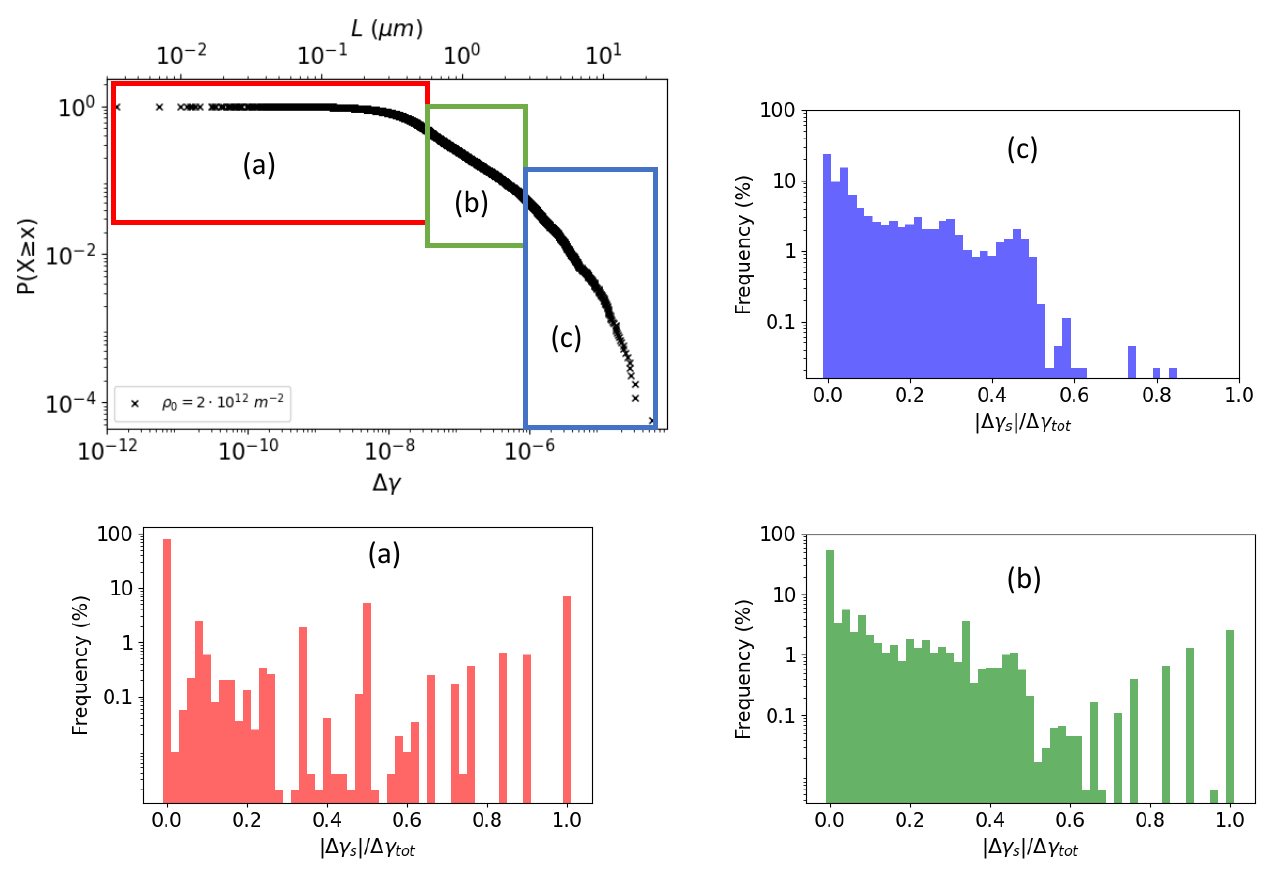}
        \caption{Histograms of avalanche contributions $c_{sa}$ for the heigh active slip systems in a [001] simulation at $\rho_0 = 10^{12}$ m$^{-2}$. Histograms are further separated into three bins depending upon the size of the avalanches shown on the top left: (a) small plastic bursts, (b) intermediate avalanches well in the power law regime and (c) largest avalanches. 
        }
        \label{fig:histoslipcontrib}
\end{figure}

\begin{figure}[h!]
        \centering
        \includegraphics[width=0.45\textwidth]{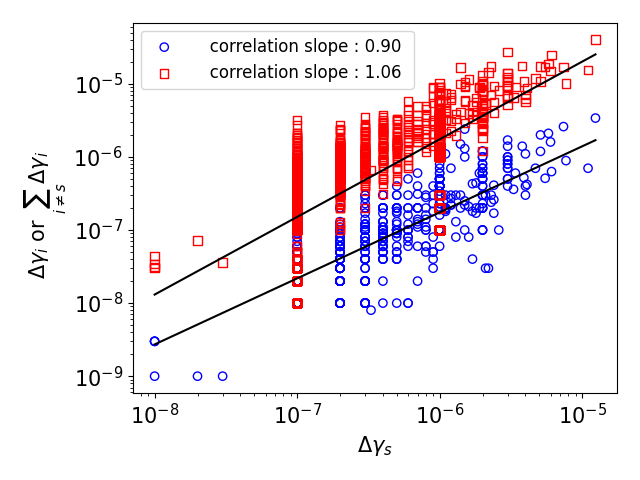}
    \caption{Correlations among slip system contributions to avalanches.  Correlation (blue) between the primary system $1/2 [\bar{1}01](1\bar{1} 1)$ and its collinear system $1/2[\bar{1}01](111)$;  correlation (red) between the primary system $1/2 [\bar{1}01](1\bar{1} 1)$ and all other systems except its collinear system $1/2[\bar{1}01](111)$. 
    } 
    \label{fig:correldgamma}
\end{figure}

From the observation of simulation sequences, we know that dislocation avalanches may be initiated by various \emph{critical events} such as: i) a single segment acting akin to a source, ii) destruction of binary or higher order junctions, iii) or snow-balling effect. Tracking these events directly as the simulated microstructure evolves is virtually impossible at the moment. {Here we propose instead a statistical analysis of the relative contributions of individual slip systems 's' though the ratio $c_{sa} = {|\Delta \gamma_s| \in a  \over \sum_j | \Delta \gamma_j| \in a}$ to an individualized avalanche 'a'. Avalanches are further categorized into three different classes (with a similar number of events) depending on their overall size: small strain bursts with $\Delta \gamma < 2 \times 10^{-7}$, intermediate avalanches in the power law regime with $2 \times 10^{-7} < \Delta \gamma < 3 \times 10^{-6}$, and the largest avalanches above $\Delta \gamma > 3\times 10^{-6}$}. 

{
Histograms of $c_{sa}$ are provided for all systems in Fig. \ref{fig:histoslipcontrib} for a [001] simulation, representative of other simulations. Typically, small plastic events correspond to quantized contributions of the different active slip systems. Very often, for $80\%$ of avalanches, not all systems contribute to small events ($c_{sa}\approx0$), for about $7\%$ of avalanches, only one system is contributing alone to the avalanche ($c_{sa}=1$). Between these extremes, $k$ active slip systems share similar contributions to $\Delta \gamma_a$ with peaks around  $c_{sa}=25\%$, $33\%$ or $50\%$ ($\approx k/n$ with $n= 8$ the total number of slip systems). This contrasts greatly with the continuous histogram obtained for the largest avalanches, where a single system cannot contribute alone ($c_{sa}=1$) and fewer cases of inactive slip systems can be found (less than 25 $\%$ of avalanches). Most of these large plastic events correspond to concomitant activity on all slip systems, with a broad peak for $0 \% < c_{sa} < 25\%$. This picture of large events where all slip systems are contributing from the beginning or when destabilized by activity in an adjoining region, matches well the general picture of avalanches. Finally, the histogram for intermediate events, in the middle of the power law regime corresponds nicely to a mixture of the two previous histograms. Besides, non-primary systems tend to contribute in average less to avalanches. Correlations among slip system activities (even when inactive according to the Schmid's law) were also recently nicely illustrated in \cite{akhondzadeh2021slip}, here we show that this correlation is rooted in the avalanche kinetics itself.}

{
Figure \ref{fig:correldgamma} further illustrates the correlations existing among slip systems, where $\Delta \gamma_s$ is plotted as function $\sum_{j\neq s} \Delta \gamma_j$. An approximate correlation is found $\Delta \gamma_i \propto \Delta \gamma_{j \neq i}^{1.06}$ marginally above linearity (with admittedly a large spreading of the data). Surprising also is the slightly different correlation found between the primary system and its cross-slip systems as $\Delta \gamma_i \propto \Delta \gamma_{CS_i}^{0.9}$ below linearity. This result is surprizing as we expect that the motion of a primary segment during an avalanche will carry with it the collinear superjogs or cross-slip segments that were already present on the initial dislocation \cite{devincre2007collinear}, thus leading to a special link between these two slip systems. This observed reduced correlation therefore suggests that cross-slip activity during an avalanche competes with that of the primary system.} 

\begin{figure}[h!]
    \centering
    \begin{subfigure}[b]{0.45\textwidth}
        \centering
        \includegraphics[width=\textwidth]{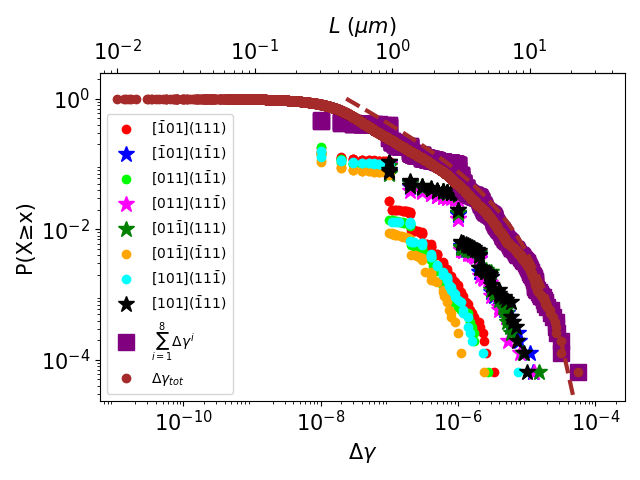}
        \caption{}
         
    \end{subfigure}
    \hfill
    \begin{subfigure}[b]{0.45\textwidth}
        \centering
        \includegraphics[width=\textwidth]{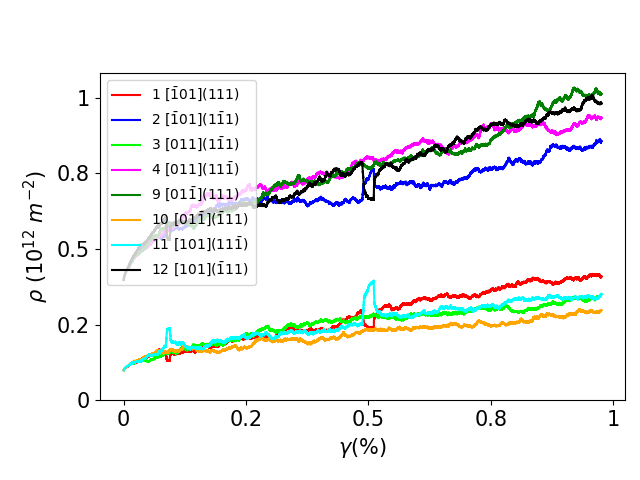}
        \caption{}
         
    \end{subfigure}
    \caption{{Statistical analysis of dislocation avalanches on a per-system basis. a) CCDF for $\Delta \gamma_i$ on each of the active slip systems are compared with the global CCDF. Fewer plastic events are found on the collinear (cross-slip) systems since the initial microstructure is made of asymmetrical prismatic loops to ensure loading stability along the [001] direction. Floating point precision is lower for $\Delta \gamma_i$ that what was used for $\Delta \gamma$. b) Evolution of the dislocation density on all slip systems during deformation along [001] direction. {Initial density is larger on four \emph{primary} systems (collinear to the other systems) so as to obtain the the hardening typical of [001] deformation and stable deformation.}}
    }
    \label{fig:CCDFpersystem}
\end{figure}

Finally, Figure \ref{fig:CCDFpersystem}.a) shows the CCDF for plastic bursts $\Delta \gamma_i$ on individual systems 'i'. The statistical signature is similar to the distribution obtained previously for the total plastic bursts $\Delta \gamma$. {The extent of the power law regime is shorter as events on individual systems are smaller,} the power law exponent is close to 1.5 for all primary and non-primary systems. Naturally, we can show that 
$p(\Delta \gamma_{tot}) = p(\sum \Delta \gamma_{i})$ (also true for the CCDF). However, $p(\Delta \gamma_{tot})\neq \sum p(\Delta \gamma_{i}) $ as correlations exist among slip system strain bursts (see above). {Ultimately, it becomes clear from the figure that $\max(\Delta \gamma_{i})$ are different for every slip systems.  Figure \ref{fig:CCDFpersystem}.b) shows that dislocation densities $\rho_i$ are not exactly the same on all primary or their collinear systems as commonly seen at the mesoscale. The obstacle density of a given  system may be defined as   $\rho^i_{obs} = \rho_{tot} - \rho_i$. For example, system 1/2 $[01\bar{1}](\bar{1}11)$ (respectively, 1/2 $[01\bar{1}]({1}11)$) experiences the largest (lowest) obstacle density, which yields to the smallest (largest) $\Delta \gamma_{i}$. Consequenctly, the upper cut off behavior does not depend upon the total dislocation density but rather upon the obstacle dislocation density seen by the considered slip system. }

\subsection{Scaling of the cutoff $\Delta \gamma_{max}$ with dislocation density and orientation}

Simulation results clearly show that the power law truncations decreases with the increase of the density of obstacles seem by avalanches. The pure power law is known to be scale free, therefore understanding the cutoffs of the power law regime becomes important to be able to define average processes of the intermittent plastic activity seen at the mesoscale and provide a rigorous derivation of plastic behaviour at the continuous macroscale. Since dislocation density evolves continuously in simulations and experiments, the cutoff $\Delta \gamma_{max}$ is ill defined, which may explain in part the wide fluctuations seen in the literature.

\begin{figure}[h!]
    \centering
    \begin{subfigure}[b]{0.45\textwidth}
        \centering
        \includegraphics[width=\textwidth]{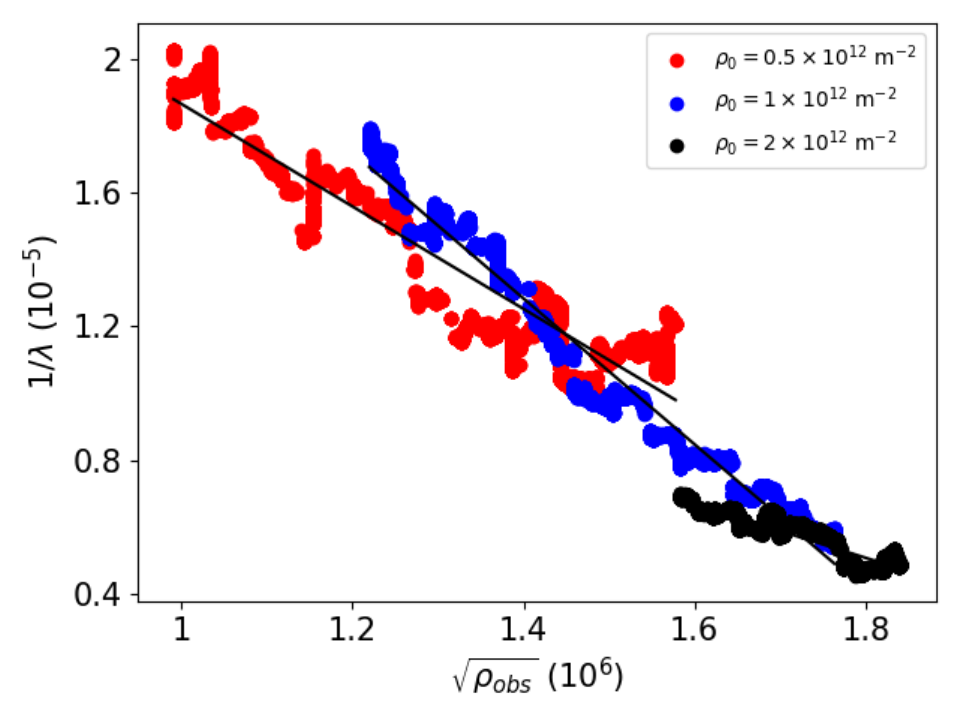}
        \caption{}
         
    \end{subfigure}
    \hfill
    \begin{subfigure}[b]{0.45\textwidth}
        \centering
        \includegraphics[width=\textwidth]{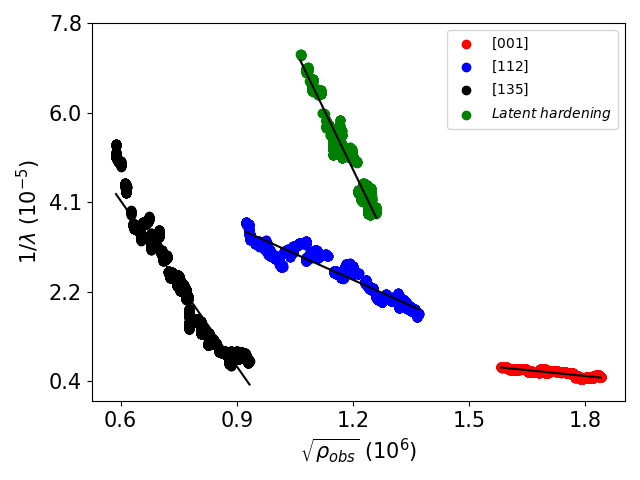}
        \caption{}
         
    \end{subfigure}
    \caption{Evolution of the parameter $\Delta \gamma_{max}=1/\lambda$ modeling the power law upper cutoff. It is calculated by means of a sliding window of 6,000 events, for which the dislocation density evolution is negligible. $\Delta \gamma_{max}$ is shown as function of $b\sqrt{\bar{\rho}_{obs}}$. a) Results for various [001] simulations at different initial dislocation density.  
    b) Anisotropy of the $\Delta \gamma_{max}$ evolution, which depends upon the loading direction. }
    \label{fig:evolxmax}
\end{figure}

Our results also provide a physical justification for the choice of function employed to model avalanche statistics. Avalanche extension is controlled by the collisions with forest obstacles, and collisions between dislocations show similarities with 1D Poisson processes \cite{sills2018dislocation}. Truncating the power law with an exponential decay is thus very appropriated: $p(\Delta \gamma) \propto \Delta \gamma^{\alpha} \exp(- \lambda \Delta \gamma)$ with $\lambda = 1/\Delta \gamma_{max}$ parameters resulting from the modeling of the DDD data ($\Delta \gamma_{max}$ is a fitting parameter and not exactly $\max(\Delta \gamma)$ from DDD data). We will thus continue using this type of function to model avalanche statistics.

To quantify the evolution of the $\Delta \gamma_{max}$ during a single simulation, we need to evaluate it over a range of events, that is sufficiently small that the dislocation density increase can be neglected, while preserving statistical significance. {We will thus fit the avalanche data using a centered running window over a finite number of plastic events. This is made possible by the large data available here. We varied the size of the running window from 700 to 10,000 events, the resulting quantitative $\Delta \gamma_{max}$ evolution is impacted by less that 10$\%$, and a window over 6,000 events is deemed sufficient for the purposes of the present study.}

{Figure \ref{fig:evolxmax} shows the evolution of the $\Delta \gamma_{max}$ parameter fitted using MLE from the DDD data. Since $\Delta \gamma_{max}$ must correspond to a shear, we propose that it scales as $\propto b\sqrt{\rho_{obs}}$. Here, we defined $\bar{\rho}_{obs}$ as a weighted average of the forest density seen by the systems 'i' participating to the avalanches 'a' under considerations: $\rho_{obs}^i = \sum_{j \neq i} \rho_{obs}^j $, and $\rho_{obs}^{\alpha} = {\sum_i c_{ia} \rho_{obs}^i} $. {The results are however not so different from the simpler average over the dislocation forest density $\rho_{obs}^i = \sum_{j \neq i} \rho_{obs}^j = \rho_{tot}-\rho^i$.} }
Starting with the [001] data under different initial dislocation density, $\Delta \gamma_{max}$ clearly decreases $\Delta \gamma_{max} \propto -C_{hkl} b \sqrt{{\bar{\rho}}_{obs}}$, with a nice overlap between the data from the different simulations. $\lambda$ evolves monotonically, which means that the avalanche statistics (or at least the cutoff) is rapidly converging to that associated with deformed dislocation microstructures.  
Finally, when considering other loading conditions, the scaling of $\Delta \gamma_{max} (\rho)$ remains negative but appears anisotropic as $C_{hkl}$ depends upon the orientation. This anisotropy is in agreement with qualitative cutoff behaviors obtained for different loading conditions in DDD simulations \cite{devincre2010scale} and experiments on single-crystalline pillars \cite{sparks2018nontrivial}.

\begin{table}[h!]
\centering
\caption{{Values of $C_{hkl}$ for different orientations.} }
\begin{tabular}{c c}
\hline
Orientation $(hkl)$ & $C_{hkl}/b$  \\
\hline
(001) & (0.059 $\pm$ 0.016) \\
(112) & 0.266 \\
(135) & 0.430 \\
Latent hardening & 0.649 \\
\hline
\end{tabular}
\end{table}

\subsection{Distribution of avalanche triggering stresses $\tau_c$}

In this last section, we show that dislocation density controls not only the scaling of avalanche statistics cutoffs but also the distribution of the critical configurations and stresses at the origin of avalanches. The onset of avalanches corresponds to a unique situation when the applied loading equals the critical stress to activate a configuration within the dislocation microstructure. This is a critical stress, since the stress increase that precedes has mostly an elastic origin (along with perhaps a slight curvature increase of dislocations), and beyond that stress, the activated dislocation segments are no longer in (static) mechanical equilibrium, the resulting effective stress leads to large instantaneous velocity \cite{fan2021strain}.

\begin{figure}[h!]
    \centering
    \begin{subfigure}[b]{0.45\textwidth}
        \centering
         \includegraphics[width=\textwidth]{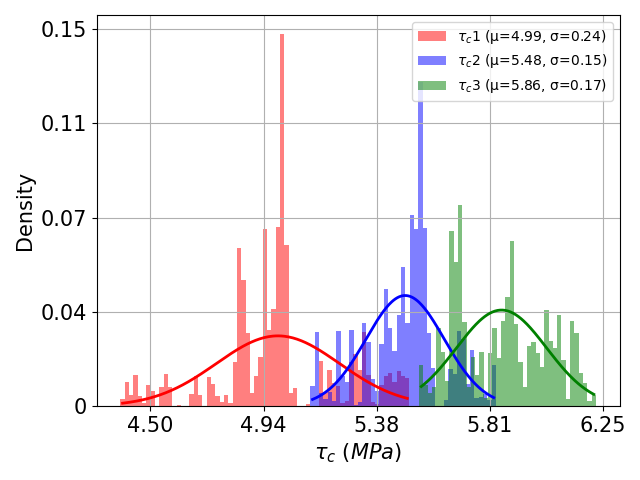}
        \caption{}
         
    \end{subfigure}
    \hfill
    \begin{subfigure}[b]{0.45\textwidth}
        \centering
        \includegraphics[width=\textwidth]{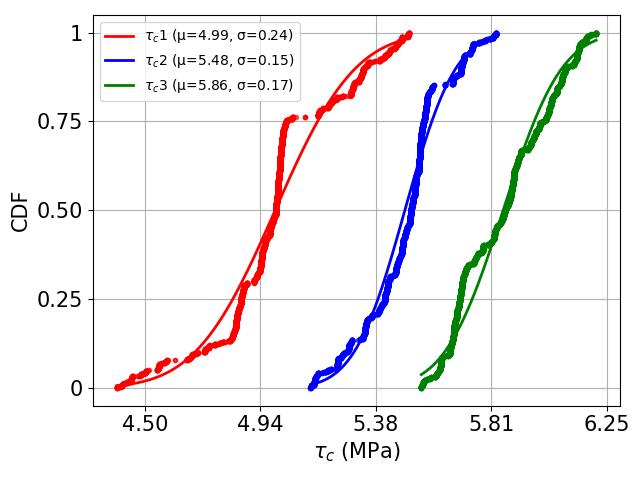}
        \caption{}
         
    \end{subfigure}
    \vfill
    \centering
    \begin{subfigure}[b]{0.45\textwidth}
        \centering
        \includegraphics[width=\textwidth]{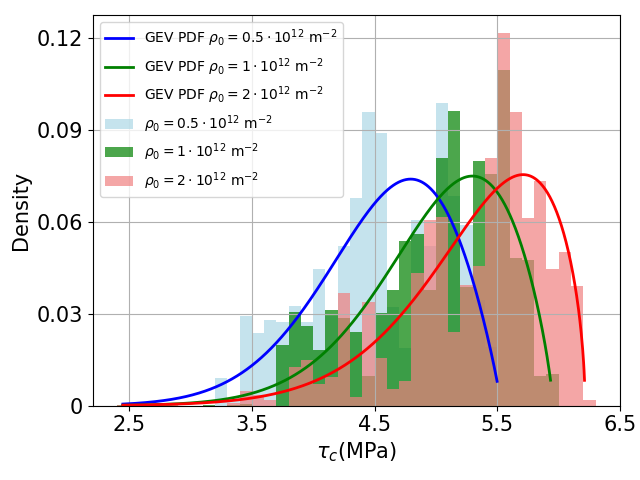}
        \caption{}
         
    \end{subfigure}
    \hfill
    \begin{subfigure}[b]{0.45\textwidth}
        \centering
        \includegraphics[width=\textwidth]{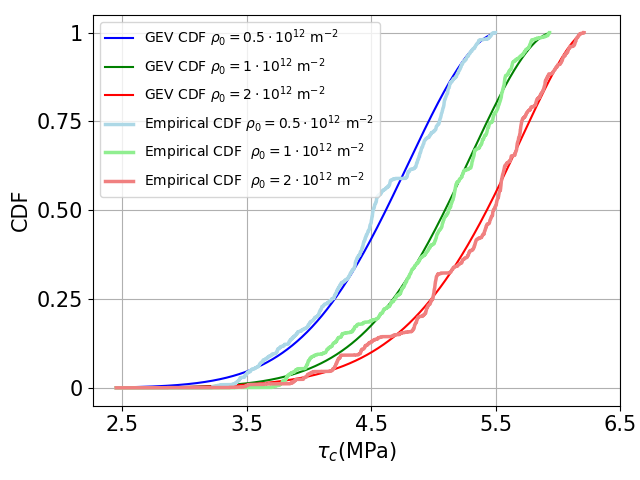}
        \caption{}
         
    \end{subfigure}
    \caption{ 
     Raw histograms of triggering stresses $\tau_c$ measured in [001] deformation simulations. (a) and (b) correspond to histograms obtained during deformation with an initial dislocation density of $\rho_0=10^{12}$ m$^{-2}$. Three ranges of deformation are considered here corresponding to different deformation states: beginning of the deformation for $\gamma < 0.46$ (in red), intermediate range for  $0.46 < \gamma < 0.73$ (in blue), and final range of the simulation for $\gamma > 0.73$ (in black). The resulting bins accounts for about the same number of events. (c), (d) regroup histograms for [001] deformation starting with different initial dislocation densities (cf. deformation curves in Fig. \ref{fig:defcurve001}). (a) and (c) are to density histograms, while (b) and (d) correspond to cumulative histograms.}
    \label{fig:rawhistotauc}
\end{figure}

Quantifying the distributions of \emph{-triggering stresses-} is crucial to understand microstructure organization and the related disorder at the origin of avalanches. This is another ingredient required to rigourously bridge the discrete nature of plastic deformation at the mesoscale to the continuous picture at the macroscale. Quantifying these distributions is however rather hard to achieve in practice as significant fluctuations may affect the analysis \cite{zaiser2006scale,berta2025identifying}. {Here, the critical stress is assumed to correspond to the last local maximum of the applied stress curve when an avalanche is triggered. We will see that these distributions evolve dynamically during deformation as critical configurations are destroyed at the beginning of an avalanche and leaving space for progressively stronger configurations when avalanches stop. }

\begin{figure}[h]
    \begin{subfigure}[b]{0.45\textwidth}
        \centering
        \includegraphics[width=\textwidth]{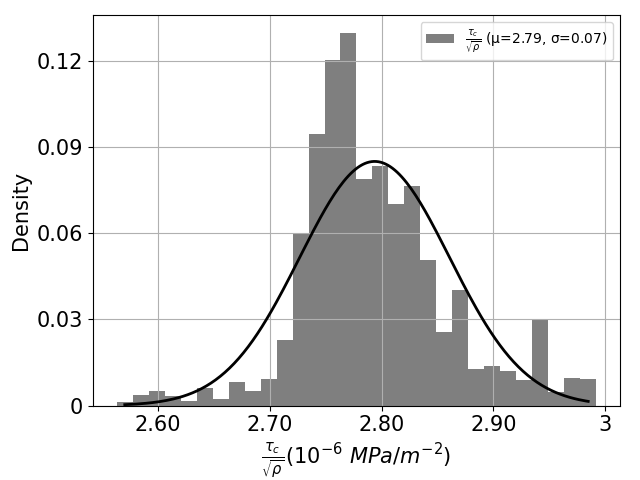}
        \caption{}
         
    \end{subfigure}
    \hfill
    \begin{subfigure}[b]{0.45\textwidth}
        \centering
        \includegraphics[width=\textwidth]{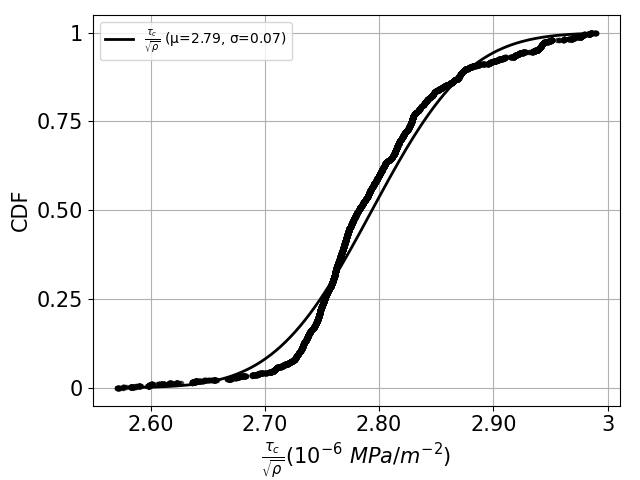}
        \caption{}
         
    \end{subfigure}
    \vfill
    \begin{subfigure}[b]{0.45\textwidth}
        \centering
        \includegraphics[width=\textwidth]{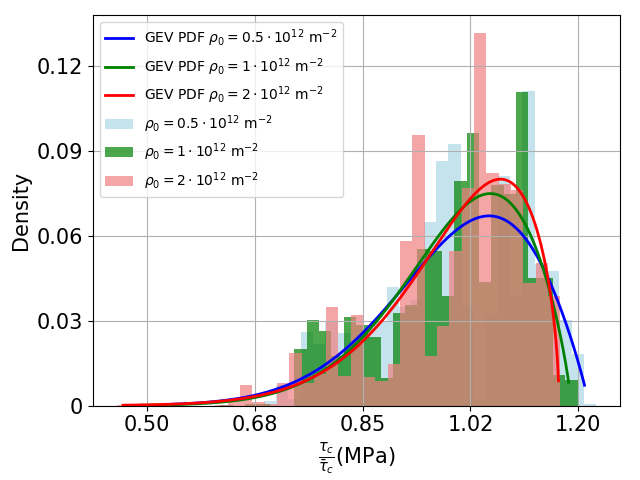}
        \caption{}
         
    \end{subfigure}
    \hfill
    \begin{subfigure}[b]{0.45\textwidth}
        \centering
        \includegraphics[width=\textwidth]{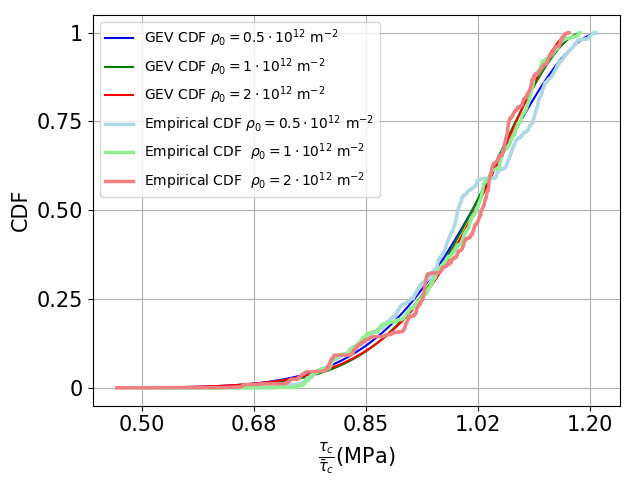}
        \caption{}
         
    \end{subfigure}
    \caption{{Histograms of triggering stresses $\tau_c$ after appropriate rescaling. The subfigures mirror the presentation of previous figure with (a) and (b) correspond to histograms  after rescaling by by $\sqrt{\rho_{obs}}$ from data obtained during deformation with an initial dislocation density of $\rho_0=10^{12}$ m$^{-2}$. (c), (d) regroup histograms for [001] deformation starting with different initial dislocation densities after rescaling by $\bar{\tau_c}$.}}
    \label{fig:rescaledhistotauc}
\end{figure}

Focusing on the [001] multislip conditions, we analyze the distributions of triggering stresses for different deformation states starting with an initial dislocation density $\rho_0=10^{12}$ m$^{-2}$. To help characterizing these distributions, we also performed a fit using a Generalized Extreme Value (GEV) probability distribution, which are displayed on the Figure \ref{fig:rescaledhistotauc}. The raw distributions are prone to large statistical fluctuations similarly to past attempts of this kind  \cite{zaiser2006scale}. However, the distributions are peaked for a $\tau_c$ values that increase with deformation, in agreement with the hardening observed. Distributions becomes less spread and more peaked as deformation progresses. This means that critical configurations associated to larger $\tau_c$ develop during deformation, and their relative contribution to the dislocation microstructure increases, while contribution of configurations associated to lower $\tau_c$ progressively disappear during deformation.

The analysis of the triggering stress distributions in Fig. \ref{fig:rawhistotauc}.(c) and (d) for different initial dislocation densities provide a similar picture. The distributions are better defined as the data for each simulation is now analyzed as a whole (with about 20,000 configurations). Histograms become less spread, more peaked, and shifted to larger stresses as the dislocation density increases. The distributions are slightly asymmetrical as they would correspond to a GEV distribution with a long tail on the low stress side.

The striking similarities among the distributions suggest that a common distribution may exist. {First following classical crystal plasticity modeling, we know that the average stress is known to follow the classical dimensionality scaling:
\begin{equation*}
\bar{\tau}^i_c = \mu b \sqrt{\sum a_{ij}\rho_{j}},
\end{equation*}
when plastic flow is controlled by dislocation reactions \cite{queyreau2009slip,saada1961interaction, madec2002dislocation, kocks2003physics,kubin2008modeling} with $a_{ij}$ the so called interaction coefficient associated to the interaction between systems i and j. It can be further simplified as $\bar{\tau}_c\approx \mu b \sqrt{\bar{a}\rho_{obs}}$ in the case of symmetrical multislip.} 

{First, Figure \ref{fig:rescaledhistotauc}.a) clearly shows that the different histograms at different strains from figure \ref{fig:rawhistotauc} can merge into a single and well defined distribution when rescaling by the obstacle density $\sqrt{\rho_{obs}}$. Next, Figure \ref{fig:rescaledhistotauc} shows that all histograms converge toward the same underlying distribution when rescaling the triggering stress by $\bar{\tau_c}$, thus including the impact of the interaction matrix $a_{ij}$. The distribution can thus be written as a Frechet distribution}: 
\begin{equation*}
     f(\tau_c) = {k \over \bar{\tau_c}} \left({ \tau_c \over \bar{\tau_c} }\right)^{k-1} \exp^{-\left({\tau_c \over \bar{\tau_c}}\right)^k}
\end{equation*}
{with an average shape parameter of $k=0.6 \pm0.05$.} \\

{This suggests that the organization of the dislocation microstructure (or at least, the critical configurations) follows the same scaling with $\sqrt{\bar{a}\rho_{obs}}$ during deformation. A comprehensive study of these distributions, their properties,
and their relationship to the dislocation microstructure as a whole is beyond the scope of this article and
will be the subject of a forthcoming article.}

\newpage
\section*{Concluding remarks}

The present paper reports a comprehensive study to assess the impact of two microstructural parameters -the dislocation density and crystal orientation- on the avalanche statistics in bulk fcc Cu deformed under a fixed strain rate. Our strain resolved simulation results evidence a well defined power law regime over several orders of magnitudes in sizes, with an exponent $\alpha \approx 1.7 \pm0.1$, which validates the range of exponents previously obtained for dislocation plasticity in 3D  \cite{weiss1997acoustic,richeton2005breakdown,sparks2018nontrivial,devincre2008dislocation}. The exponent is unaffected by dislocation density and loading conditions suggesting that these parameters have no impact over the associated class of universality. These parameters can thus be ruled out as an explanation for the variation in power law exponent observed in the literature. For a given material, the power law exponent is only controlled by the imposed strain rate \cite{sparks2018nontrivial,kurunczi2023avalanches} as will be shown in our forthcoming paper \cite{aissaoui2025connecting} . We can therefore conclude that the exponents obtained in the case of polycrystals made up of aggregates of grains of potentially different dislocation densities and orientations are well defined as long as the strain rate inside the grains where avalanches are active is the same. This condition is generally considered to be true at small strain, but may not apply in latter stages of deformation.

{ We initially expected that the relative importance of short- and long-range interaction contributions would strongly influence the avalanche statistics when changing dislocation density and microstructure.}  For the former, a small effect was expected, as the self-stress scaling with dislocation density (through $l\propto 1/\sqrt{\rho}$) while not formally the same, is not so far from the scaling of long-range elastic contributions with density. However, a stronger impact was expected from the different loadings and deformation paths considered here. {Sequences of simulated avalanches have illustrated this in  Fig. \ref{fig:supp-avalan}}. Single glide deformation is typically associated to strong multipolar distant interactions within the entanglements formed, with few reactions associated to collision with collinear superjogs \cite{devincre2007collinear}. This contrasts with [112] and [001] multislip conditions where plastic flow proceeds with a large number contact reactions and are associated to the formation of planar or cellular microstructures. Finally, the latent hardening simulation is very different from the previous conditions as the part of the  microstructure corresponding to the inactive forest systems is imposed from the start, and should not evolve significantly. {This latter condition is very interesting as the organization of the primary system is related to the random distribution of forest dislocations inserted initially and not to the \emph{self-organization} of the microstructure when all systems are simultaneously operating and interacting as for [112] or [001] loading directions. Despite these differences, the resulting exponents in the power law regime are identical and this is not entirely understood by the authors.}

Our data shows that the power law truncation can be modeled as: 
\begin{equation}
\Delta \gamma_{max} = D_{hkl}-C_{hkl}b\sqrt{\rho_{obs}}.
\end{equation}
This result is in qualitative agreement with the scaling suggested by Csikor and collaborators \cite{csikor2007dislocation} from simulation and experimental data as: 
\begin{equation}
\Delta \gamma_{max} \propto {b E\over D(\theta +M)},
\end{equation}
where $E$ is the Young's modulus, $\theta$ a hardening strain rate, and $M$ the combined stiffness of the both the specimen and the testing machine (here $M \approx E$). {The length $D$ is microstructure dependent ($\neq D_{hkl}$ our parameter) and can be used to capture size effects as the one we obtained here.  Dislocation microstructures follow the \emph{similitude principle} and typically scale as $D\propto 1 / \sqrt{\rho}$, which is thus in agreement with our findings. It is easy to see that $C_{hkl}$ should thus be $\propto E/D(\theta + M)\sqrt{\rho_{obs}}$ And this is only partially in agreement with our results. When comparing results for [135], [112] and [001] loading directions, the hardening rate $\theta$ is decreasing corresponding to smaller $\Delta \gamma_{max}$, which is in qualitative agreement with our measurements of $C_{hkl}$ however the quantitative scaling does not seem to follow $E/D(\theta + M)$ as both $\theta$ and $D$ are orientation dependent, we cannot therefore separate their impact. Besides, the hardening rates simulated for the [135] deformation and the latent hardening are naturally barely noticeable for a few percents of deformation, however our $C_{135}$ and $C_{lat.hard.}$ show very different values. Additional study is certainly necessary for a definite scaling of $\Delta \gamma_{max}$ with $\theta$. For now, it is sufficient to state that $C_{hkl}$ is orientation dependent.}

The dependence of the cutoff with dislocation density has several important consequences. First, most of the existing data from simulations or experiments has been obtained during deformation, and depending on the system, dislocation density may have largely increased. {Avalanche statistics, which are generally established from entire experiments, are therefore a superposition of distributions with an changing cutoff behavior. This could obviously make it difficult to define the distribution cutoff, but if the event ranges are limited or the statistics are low, this could lead to some degree of rounding in the power law regime, or even skew the measurement of the power law exponent.} Second, the avalanche statistics is therefore not only a function of $\Delta \gamma$ but also  of the dislocation density $\rho$:
\begin{equation}
    p(\Delta \gamma, \rho) = {1 \over C(\rho)} \Delta \gamma^{\alpha} \exp(- \Delta \gamma / \Delta \gamma_{max}(\rho))
\end{equation}
{As a first step, we quantified the evolution of $\Delta \gamma_{max}(\rho)$ in a deterministic manner, and this will have an important consequence in modifying the normalization constant $C(\rho)$, which evolves with deformation.}

{Finally, in some experimental works \cite{weiss2019plastic}, acoustic emission signals were separated into a continuous AE signal associated to a Gaussian distribution of amplitudes for which plasticity is mild, and discrete AE signal associated to a power law regime for which plasticity is wild.} As deformation proceeds, the wild part of the signal diminishes. Our quantitative results confirm the interpretation from the authors \cite{weiss2019plastic}, as the discrete part of plastic flow that can be detected by AE, (that is above the background threshold and below the powerl cutoff) has to decrease.

\newpage
\section*{Supplementary Material}
\subsection*{Avalanche analysis and modeling}
Dislocation avalanche amplitudes \cite{devincre2008dislocation,miguel2001intermittent} typically follow a power law distribution of the form \( p(x) = C x^{-\alpha} \). {Power law analysis may be delicate when seeking for quantitative results, here, we mostly reprise the approach from \cite{clauset2009power}}. We will employ a truncated PWL distribution with an exponential decay:  \( p(x) = Cx^{-\alpha} e^{-\lambda x} \). two parameters need to be defined:  $\alpha$ and $\lambda$ \cite{clauset2009power}.

The complementary cumulative distribution function (CCDF), provides a complementary view of a variable distributed according to a power law, denoted as \( P(x) \), which for the continuous case is defined as \( P(x) = \Pr(X \geq x) \):
\[
P(x) = \int_{x}^{\infty} p(x') \, dx' = \left( \frac{x_{\min}}{x} \right)^{\alpha - 1}. \tag{4}
\]
Generally, the visual shape of the CDF is more robust than that of the probability density function (pdf) against fluctuations due to finite sample sizes, especially in the tail of the distribution.

Considering the continuous power-law distribution:
\[
p(x) = (\alpha - 1) x_{\min}^{\alpha - 1} x^{-\alpha}, \tag{5}
\]
where $\alpha$ is the scaling parameter and $x_{\min}$ is the minimum value from which the power-law behavior is observed. Given a dataset containing $n$ observations $x_i \geq x_{\min}$, we would like to determine the value of $\alpha$ for the power-law model that is most likely to have generated our data. The probability that the data are drawn from the model is proportional to
\[
p(x \mid \alpha) = \left( \frac{\alpha - 1}{x_{\min}} \right)^n \prod_{i=1}^n \left( \frac{x_i}{x_{\min}} \right)^{-\alpha}. \tag{6}
\]
This probability is the likelihood of the data given the model \cite{clauset2009power}. The data are most likely to have been generated by the model with the scaling parameter $\alpha$ that maximizes this function. In practice, we often work with the logarithm $L$ of the likelihood, which reaches its maximum at the same point:
\[
L = \ln p(x \mid \alpha) = n \ln(\alpha - 1) - n \ln x_{\min} - \alpha \sum_{i=1}^n \ln \left( \frac{x_i}{x_{\min}} \right)
\]
\[
= n \ln(\alpha - 1) - n \ln x_{\min} - \alpha \sum_{i=1}^n \ln \left( \frac{x_i}{x_{\min}} \right). \tag{7}
\]
By setting $\frac{\partial L}{\partial \alpha} = 0$ and solving for $\alpha$, we obtain the maximum likelihood estimate (MLE) for the scaling parameter:
\[
\hat{\alpha} = 1 + n \left[ \sum_{i=1}^n \ln \left( \frac{x_i}{x_{\min}} \right) \right]^{-1}. \tag{9}
\]

\subsection*{Dislocation microstructures and avalanche sequences}

\begin{figure}[h!]
    \centering
    \captionsetup{font=small} 

    \begin{subfigure}[b]{0.48\textwidth}
        \centering
        \includegraphics[width=0.9\textwidth]{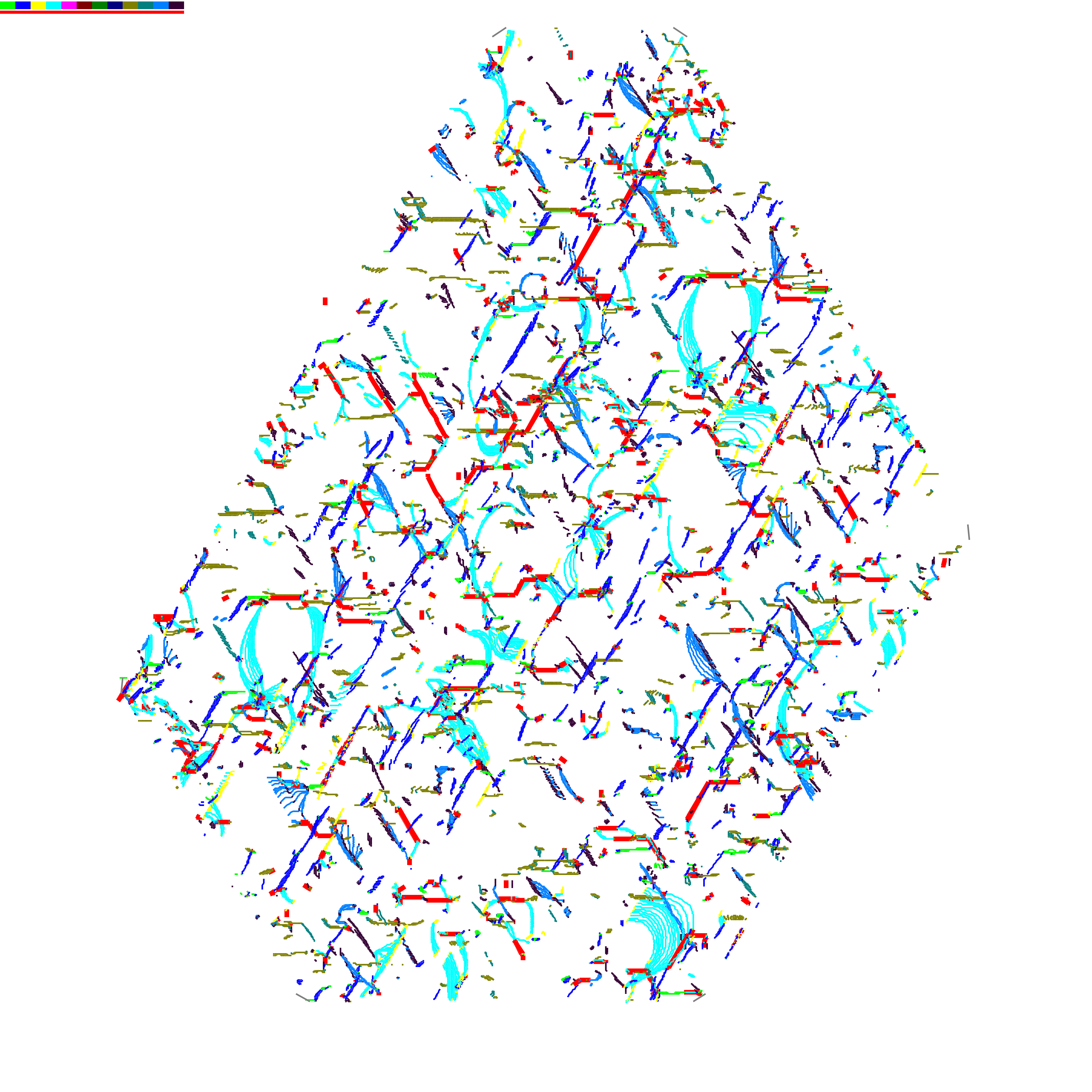}
        \begin{tikzpicture}[overlay, font=\scriptsize]
            \node at (-7.5, 4) {
                \begin{tikzpicture}[scale=0.5]
                    \draw[thick] (0,0) circle [radius=0.2] node[anchor=east] {$[11\Bar{1}]$};
                    \draw[thick] (-0.1,-0.1) -- (0.1,0.1); 
                    \draw[thick] (-0.1,0.1) -- (0.1,-0.1);
                    \draw[->, thick] (0,0) -- (1,1) node[anchor=south west] {$[2\Bar{1}1]$};
                    \draw[->, thick] (0,0) -- (1,-1) node[anchor=north west] {$[011]$};
                \end{tikzpicture}
            };
            \draw[thick] (0,0) -- (1,0) node[midway, above] {15 $\mu$m};
        \end{tikzpicture}
        \caption{[001]}
    \end{subfigure}
    \hfill
    \begin{subfigure}[b]{0.48\textwidth}
        \centering
        \includegraphics[width=0.9\textwidth]{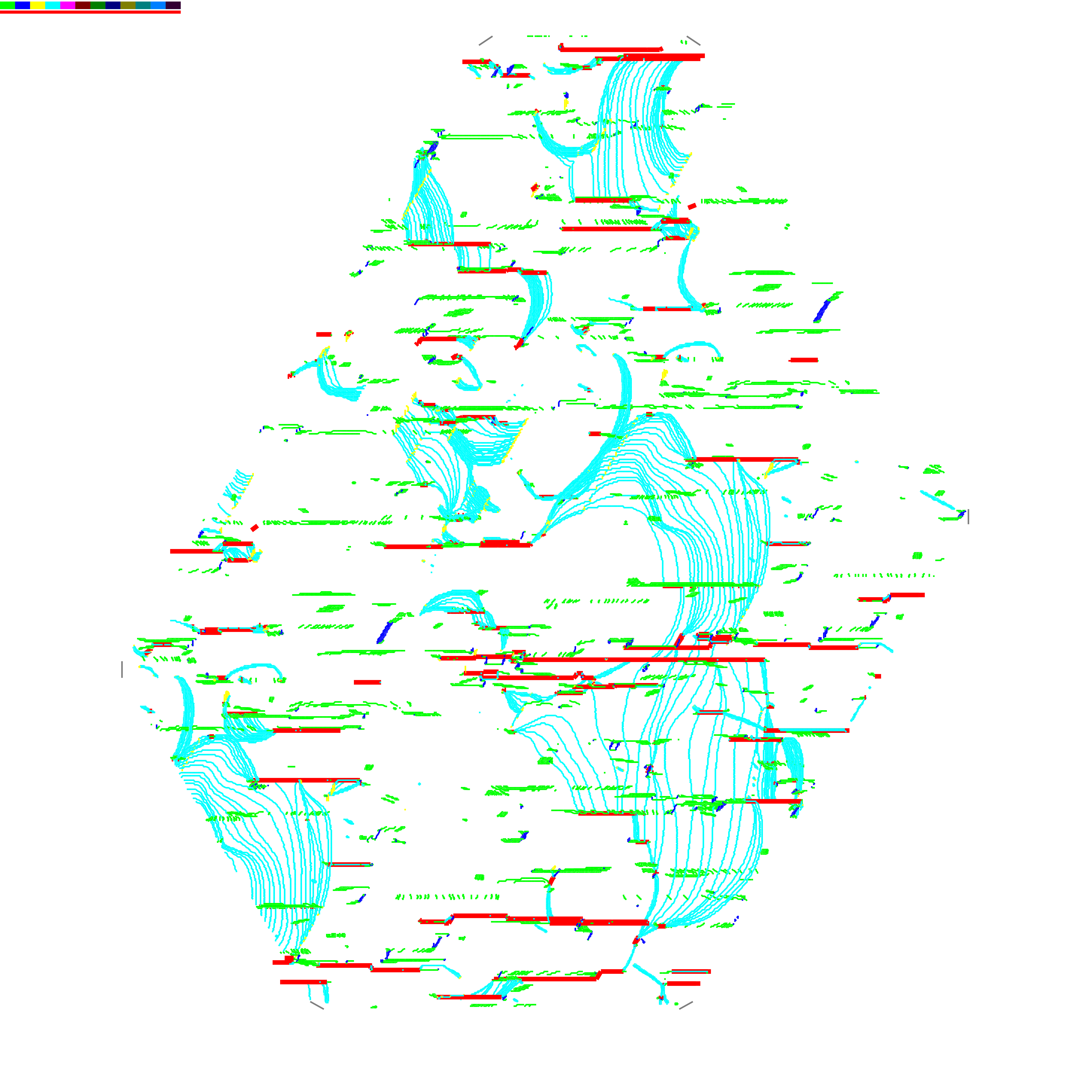}
        \begin{tikzpicture}[overlay, font=\scriptsize]
            \node at (-7.5, 4) {
                \begin{tikzpicture}[scale=0.5]
                    \draw[thick] (0,0) circle [radius=0.2] node[anchor=east] {$[11\Bar{1}]$};
                    \draw[thick] (-0.1,-0.1) -- (0.1,0.1); 
                    \draw[thick] (-0.1,0.1) -- (0.1,-0.1);
                    \draw[->, thick] (0,0) -- (1,1) node[anchor=south west] {$[2\Bar{1}1]$};
                    \draw[->, thick] (0,0) -- (1,-1) node[anchor=north west] {$[011]$};
                \end{tikzpicture}
            };
            \draw[thick] (0,0) -- (1,0) node[midway, above] {15 $\mu$m};
        \end{tikzpicture}
        \caption{[112]}
    \end{subfigure}

    \vspace{0.5cm} 

    \begin{subfigure}[b]{0.48\textwidth}
        \centering
        \includegraphics[width=0.9\textwidth]{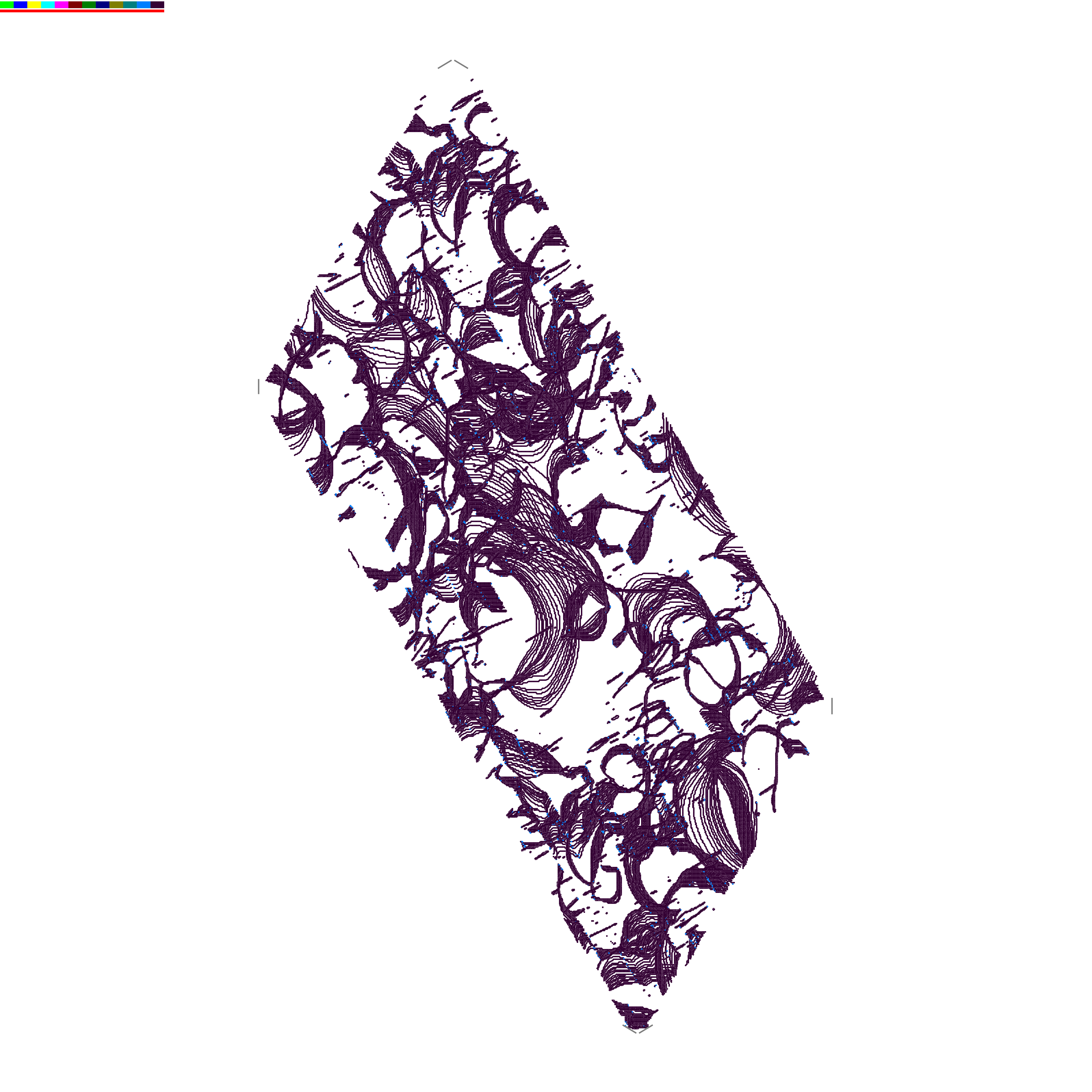}
        \begin{tikzpicture}[overlay, font=\scriptsize]
            \node at (-7.5, 4) {
                \begin{tikzpicture}[scale=0.5]
                    \draw[thick] (0,0) circle [radius=0.2] node[anchor=east] {$[\Bar{1}11]$};
                    \draw[thick] (-0.1,-0.1) -- (0.1,0.1); 
                    \draw[thick] (-0.1,0.1) -- (0.1,-0.1);
                    \draw[->, thick] (0,0) -- (1,1) node[anchor=south west] {$[101]$};
                    \draw[->, thick] (0,0) -- (1,-1) node[anchor=north west] {$[12\Bar{1}]$};
                \end{tikzpicture}
            };
            \draw[thick] (0,0) -- (1,0) node[midway, above] {15 $\mu$m};
        \end{tikzpicture}
        \caption{[135]}
    \end{subfigure}
    \hfill
    \begin{subfigure}[b]{0.48\textwidth}
        \centering
        \includegraphics[width=0.9\textwidth]{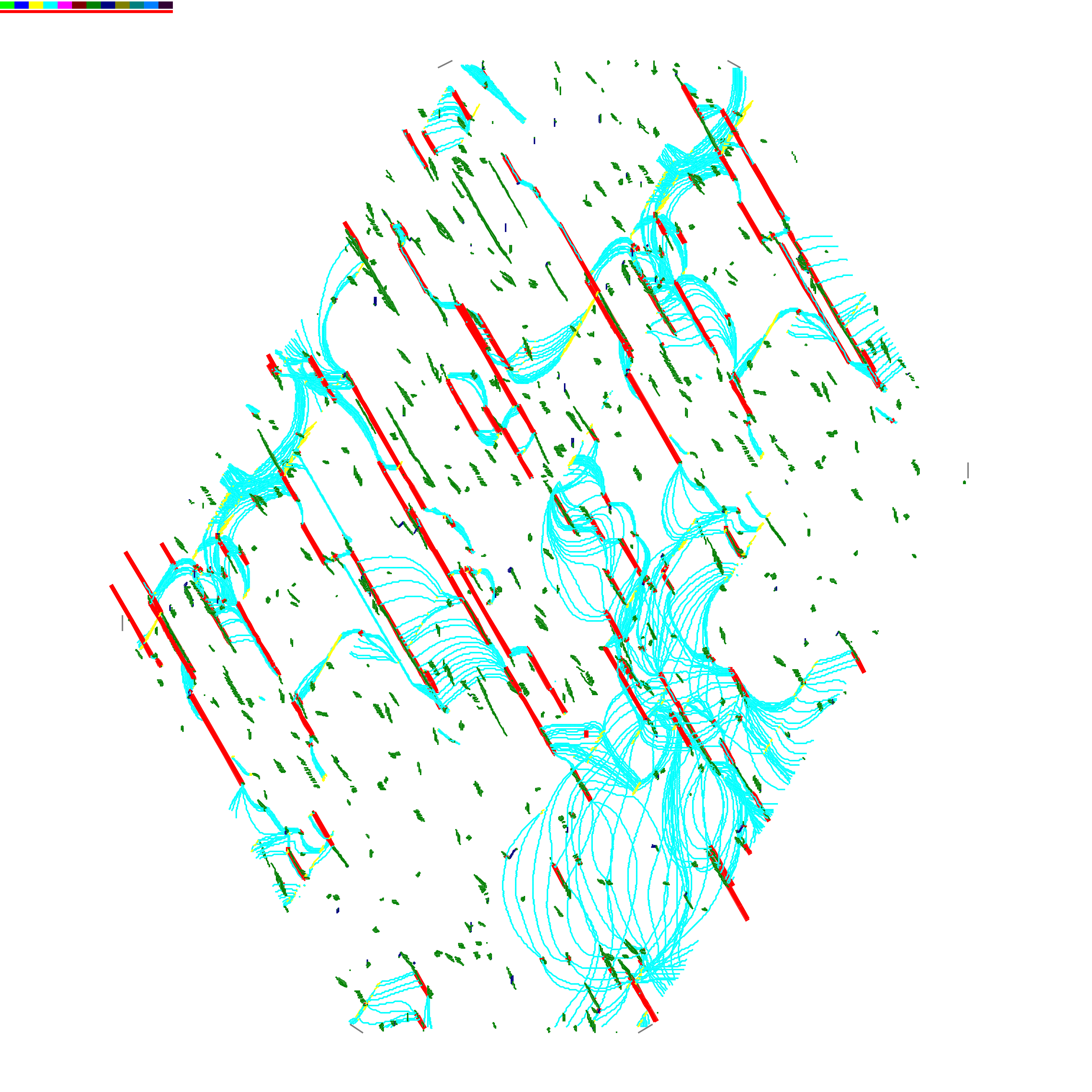}
        \begin{tikzpicture}[overlay, font=\scriptsize]
            \node at (-7.5, 4) {
                \begin{tikzpicture}[scale=0.5]
                    \draw[thick] (0,0) circle [radius=0.2] node[anchor=east] {$[11\Bar{1}]$};
                    \draw[thick] (-0.1,-0.1) -- (0.1,0.1); 
                    \draw[thick] (-0.1,0.1) -- (0.1,-0.1);
                    \draw[->, thick] (0,0) -- (1,1) node[anchor=south west] {$[2\Bar{1}1]$};
                    \draw[->, thick] (0,0) -- (1,-1) node[anchor=north west] {$[011]$};
                \end{tikzpicture}
            };
            \draw[thick] (0,0) -- (1,0) node[midway, above] {15 $\mu$m};
        \end{tikzpicture}
        \caption{latent hardening}
    \end{subfigure}

    \caption{{Evolution of the microstructure during some of the largest events: (a), (b), and (d) correspond to successive dislocation events occurring in the slip system $s = [011](11\Bar{1})$ of a crystal deforming along the directions $[001]$ and $[\bar{1}12]$, as well as latent hardening. Superimposed configurations, taken at constant time intervals of 2~ns, are shown in a thin foil of 0.2~$\mu$m thickness containing the active slip planes $(11\Bar{1})$. The active system (blue lines) forms junctions (red lines) with the forest slip systems (other colors). Subfig. (c) corresponds to successive dislocation events occurring in the slip system $s = [101](\Bar{1}11)$ of a crystal deforming along the direction $[\bar{1}35]$.}}
    \label{fig:supp-avalan}
\end{figure}

{Here, we detail how initial configurations are designed to model the plastic response of the various crystal orientations and illustrate the type of microstructures formed.  The initial conofiguration consists of a random distribution of prismatic dislocation loops. Primary edge segments are connected with collinear edge segments. Primary length is chosen to be $L_p \approx 4 \Lambda$ with $\Lambda = 1/\sqrt{\rho_0}$ the average dislocation spacing at total density $\rho_0$, and collinear length $L_{c} \approx \Lambda$, leading to steady state cross-slip and collinear activity for our material parameters (confirmed in Fig. 6.(b)). Figures \ref{fig:supp-avalan} (a), (b), (c), and (d) illustrate avalanche sequences (around some of the large avalanches) obtained by superimposing microstructure images at time intervals of $\Delta t = 2$ ns. Thinfoils are orientated along plane $(11\bar{1})$ for fig. (a, b, and c), and $(\bar{1}11)$ for fig. (d). The avalanche propagation modes differ noticeably as a function of the loading conditions. }

{\color{blue}

\begin{itemize}

\item \textbf{$[\bar{1}35]$}: this classical loading orientation corresponds to the single-glide of system: $[101] (\bar{1}11)$ named as A3 according to the Schmid and Boas convention. Self-interactions dominate the dislocation evolution, leading to a tangled microstructure characterized by strong multipolar long-range interactions and some collinear interactions with the corresponding collinear system and cros-slip segments. The avalanches observed for the $[\bar{1}35]$  orientation in Fig. (c) exhibits a very distinct behavior, in the absence of locking junctions: multipolar interactions within the same slip system, as well as collinear interactions with cross-slip segments, are highly mobile, almost the entire microstructure is in motion.

\item $[001]$: Loading along the $[001]$ orientation typically activate four (out of eight possible) slip systems. The primary slip systems are the following 
$[011] (\bar{1}\bar{1}1)$, $[101] (\bar{1}11)$, $[0\bar{1}1] (111)$, and $[\bar{1}01] (1\bar{1}1)$, which correspond to C1, A3, B2, and D4 in Schmid and Boas convention. This choice of slip systems ensures stable deformation under $[001]$ loading and reproduces the appropriate strain hardening behavior. This loading condition typically leads to various types of binary and higher order junctions and the formation of cellular microstructures. In this multiple slip condition, avalanches in Fig. (a) are distributed over several active sites (in agreement with Fig. \ref{fig:histoslipcontrib}) for the considered system and appear strongly constrained by the presence of numerous obstacle dislocations and junctions. Concomitant plastic activity on other systems traversing the thinfoil is also visible.

\item \textbf{$[\bar{1}12]$}: two slip systems are activated: $[\bar{1}01] (111)$ and $[011] (\bar{1}\bar{1}1)$, corresponding to C1 and B4 in Schmid and Boas notation. In this case, two types of interactions dominate: self-interactions (including interactions with collinear systems) and the formation of Lomer junctions. Deformation under this loading condition generally produces planar dislocation microstructures characterized by walls aligned along the unique junction direction. Fig. (b) shows that dislocation propagation localizes within channels depleted of dislocations, with facilitated motion along the Lomer junctions (in red); wall-like structures begin to form. 

\item Latent hardening: The selected primary slip system is 
$[011] (11\bar{1})$ with an initial dislocation density $\rho_p = 10^{12}\,m^{-2}$. Primary dislocations interact with a forest dislocation density defined as $\rho_f = 5\rho_p$. The forest dislocations consist of a random distribution of dislocation segments of length $2\rho_f^{-1/2}$, with randomly distributed characters. These segments are positioned such that their interaction with the primary slip system produces a single type of reaction: the formation of Lomer junctions with the forest system $[\bar{1}01] (111)$. 

This simulation differs significantly from the previous ones because the forest microstructure remains mostly fixed and is no longer solely the result of dislocation self-organization during plastic deformation. The dynamic sequence in Fig. (d) displays intermediate characteristics: dislocations in the primary system propagate in a relatively \emph{fluid} manner, similar to the $[\bar{1}35]$ case, with temporary pinning points associated with fixed junctions (in red).

\end{itemize}}

\newpage
\subsection*{Tesselation of the extended slip planes}

\begin{figure}[ht]
    \centering
    \begin{subfigure}[b]{0.45\textwidth}
        \centering
        \includegraphics[width=\textwidth]{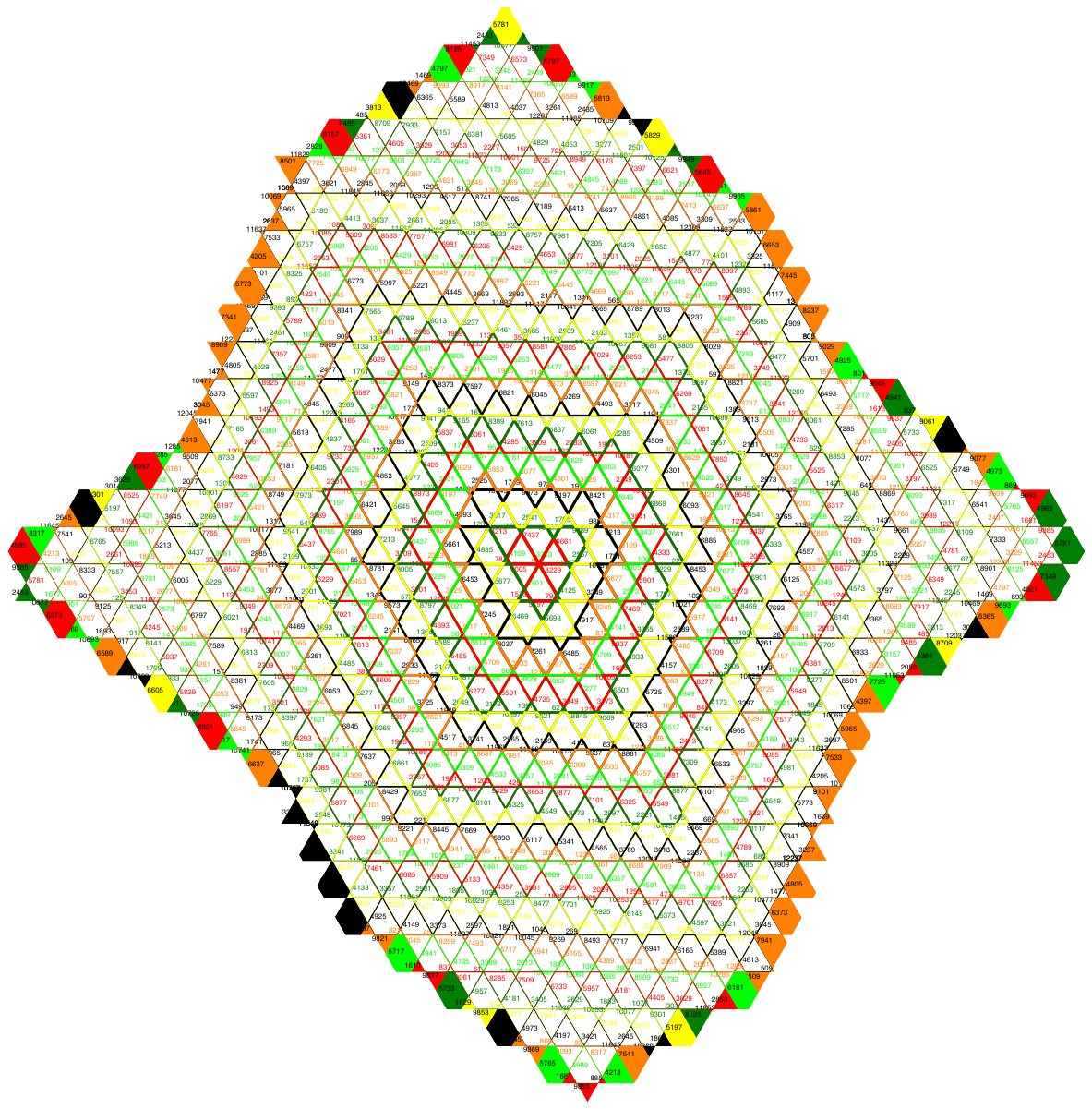}
    \end{subfigure}
    \hfill
    \begin{subfigure}[b]{0.45\textwidth}
        \centering
        \includegraphics[width=\textwidth]{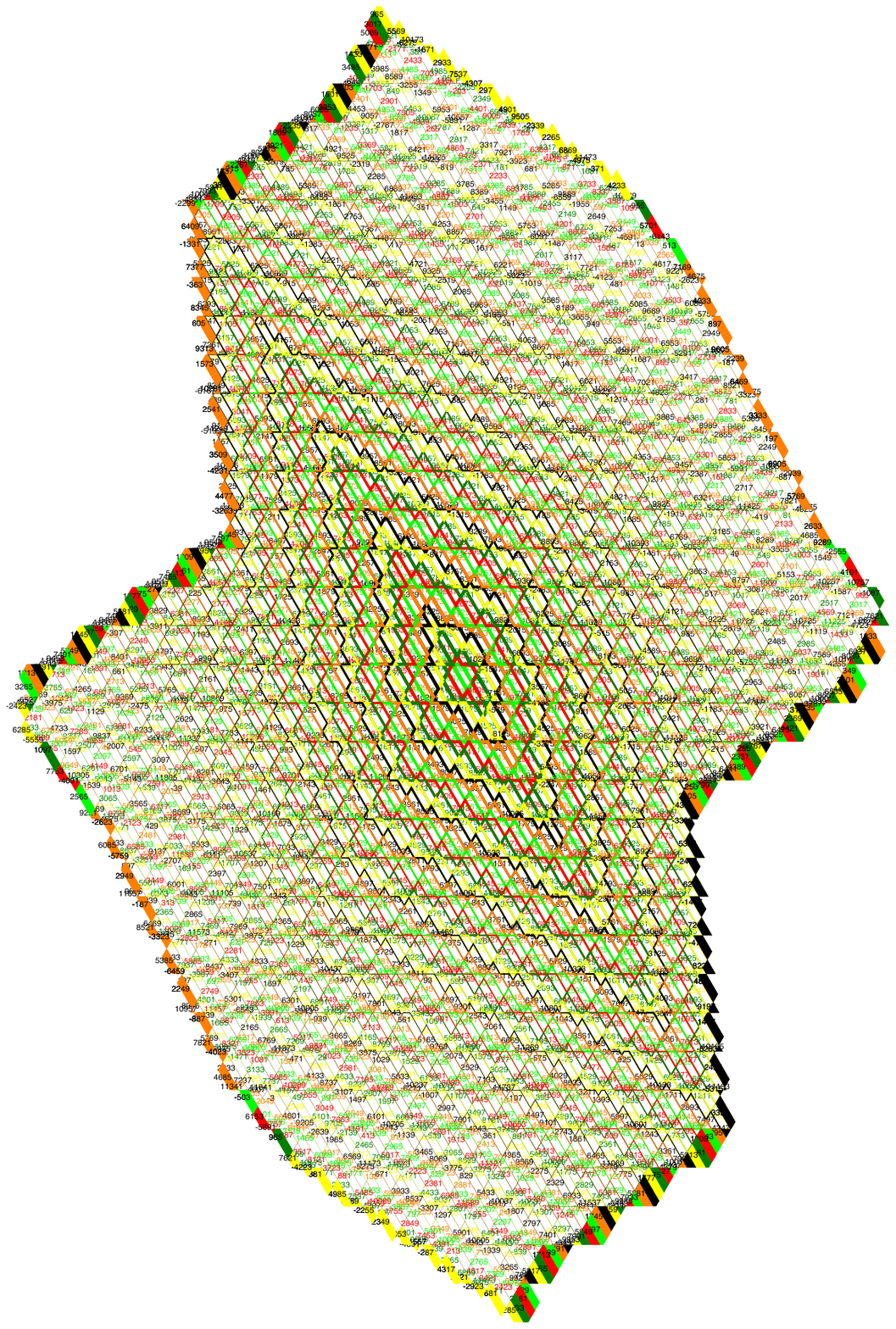}
    \end{subfigure}
    \caption{Mapping of the extended slip plane for the primary system used in a) multislip and latent hardening simulations conditions, b) [135] single slip conditions. }
    \label{fig:extendedplane}
\end{figure}

The use of periodic boundary conditions may lead to known artifacts, the main one being the self interaction of an expanding loop with itself \cite{gladwell_use_2004}. Preventing this artifact is key in capturing the natural extension of the largest avalanches and the PWL upper cutoff. Self annihilation can be delayed by carefully choosing an orthorhombic simulation box geometry. The regions of the crystals explored by an expanding loop belonging to a given slip plane are mapped using the methodology in \cite{gladwell_use_2004}. The \emph{extended slip plane} corresponds to the maximum extension possible for a slip system. Extended slip planes for the primary slip systems are given in the figure \ref{fig:extendedplane} for the box geometry optimized for i) [001], [112] and latent hardening simulations; ii) for [135] single glide conditions. The optimized geometries are a trade off between finding the largest extended plane area and the anisotropy of the geometry. We designed a larger extended slip plane for the specific [135] loading direction as dislocation mean free path is much larger in single slip condition. The paving include 1541 and 4625 polygons; the shortest radius of self-annihilation is 65 and 95 $\mu$m for geometry i) and ii), respectively.

\subsection*{Deformation curves for various loading conditions}

\begin{figure}[h]
    \centering
    \begin{subfigure}[b]{0.45\textwidth}
        \centering
        \includegraphics[width=\textwidth]{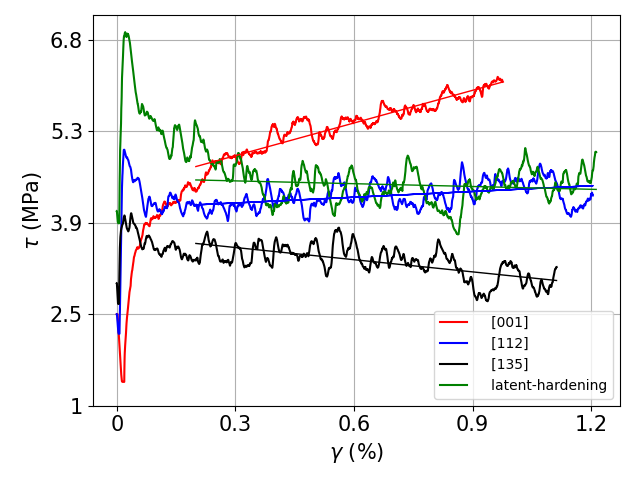}

        \label{fig:figure1}
    \end{subfigure}
    \hfill
    \begin{subfigure}[b]{0.45\textwidth}
        \centering
        \includegraphics[width=\textwidth]{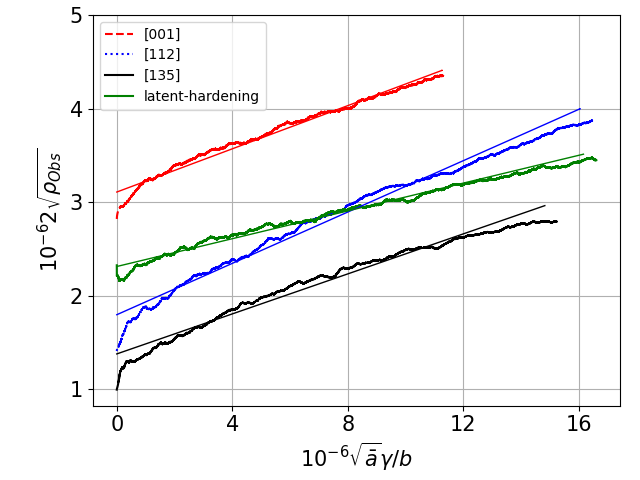}

        \label{fig:figure2}
    \end{subfigure}
    \caption{Plastic deformation simulated for various loading conditions. a) Deformation curves. b) Evolution of the total dislocation density as a function of plastic deformation. }
    \label{fig:defcurvesorientation}
\end{figure}

We considered various loading conditions to probe the anisotropy of plastic deformation, from the [135] single glide condition, [112] double slip, to [001] stable multislip condition. Latent hardening, where the forest slip system is inactive, is also considered with a forest density of $1 \times  10^{12} \ m^{-2}$. The initial dislocation density on active slip systems is taken as $5  \times 10^{11} \ m^{-2}$. The deformation curves are given in Figure \ref{fig:defcurvesorientation}.a). Multiplication peaks are visible for the [135] and [112] deformation curves as the initial dislocation density is not sufficient to produce the applied strain rate of 50 s${^{-1}}$, initially. A larger peak is visible at the beginning of  the latent hardening deformation but it corresponds to the stress to activate the smaller prismatic loops employed in this simulation \cite{queyreau2010orowan}. Past these initial peaks, we recover the expected hardening rate for these loading directions. [135] deformation exhibits a hardening close to zero, in agreement with the typical value of $\mu/3000$ seen in experiments, which is is too small to be seen over the range of deformation reached by DDD (A slight softening may even be possible and could be due to the microstructure development in the shape of entanglements). No hardening is expected as well for the latent hardening simulation as the forest density is fixed. Then, [112] simulation exhibit a linear hardening of the order of $\mu/300$ in agreement with macroscale experiments. Finally, [001] exhibits a linear hardening of about $\mu/150$ in agreement with initial slope seen in experimental data and the limited amount of dynamical recovery.

Figure \ref{fig:defcurvesorientation}.b) displays the evolution of the  the total density $\rho$ with plastic deformation using representation proposed in \cite{devincre2010scale}. $\sqrt{\rho}$ scales with $\sqrt{\bar{a}} \gamma /b$ when plastic deformation is balanced among active slip systems and dynamic recovery is negligible as here.

\subsection*{Stress resolved avalanche statistics}

\begin{figure}[h]
   \begin{subfigure}[b]{0.45\textwidth}
        \centering
        \includegraphics[width=\textwidth]{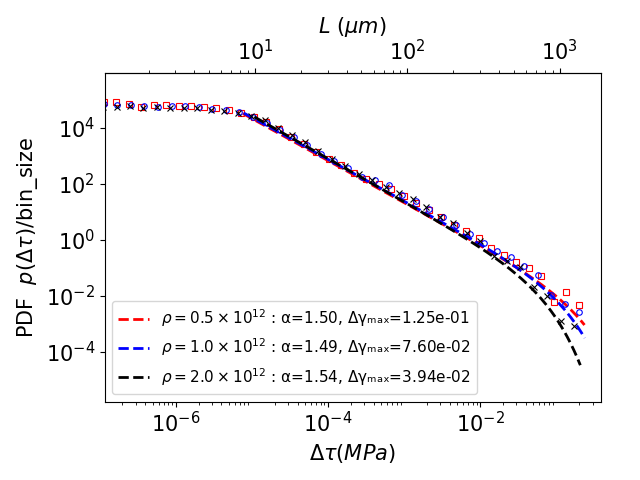}
        \caption{}
         
    \end{subfigure}
    \hfill
    \begin{subfigure}[b]{0.45\textwidth}
        \centering
        \includegraphics[width=\textwidth]{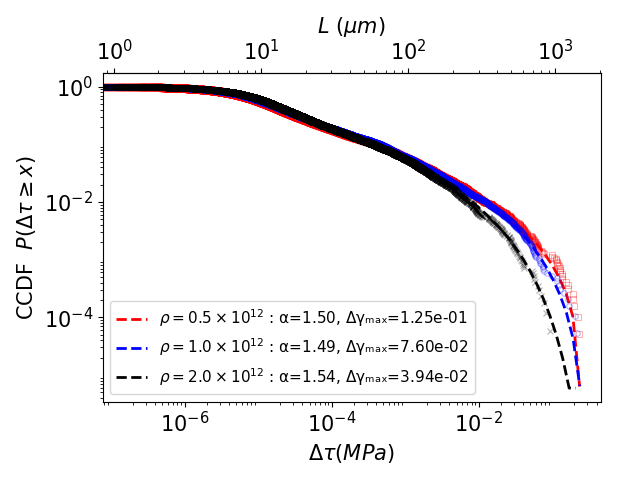}
        \caption{}
         
    \end{subfigure}
    \caption{ Impact of the initial dislocation density upon the stress resolved avalanche statistics. (a) Probability density function (PDF) of stress drops  \(\Delta\tau\). (b) Complementary Cumulative Distribution Function (CCDF) of stress-drop \(\Delta\tau\) }
    \label{fig:stressresolved_density}
\end{figure}

\begin{figure}[h!]
    \begin{subfigure}[b]{0.45\textwidth}
        \centering
        \includegraphics[width=\textwidth]{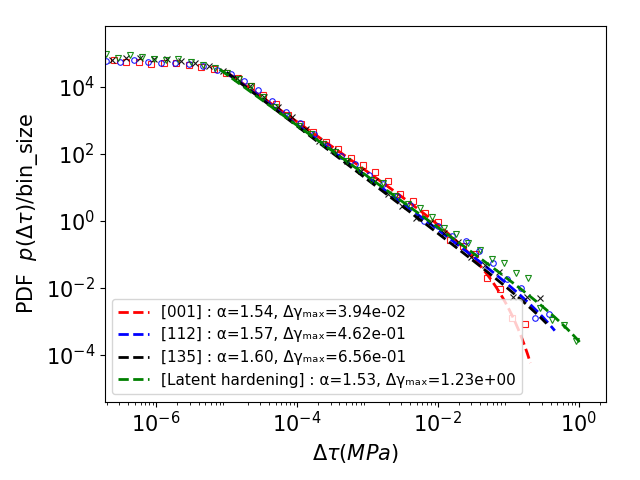}
        \caption{}
         
    \end{subfigure}
    \hfill
    \begin{subfigure}[b]{0.45\textwidth}
        \centering
        \includegraphics[width=\textwidth]{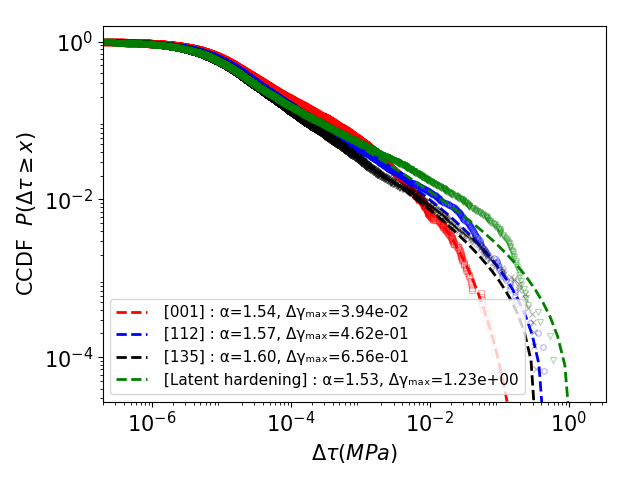}
        \caption{}
         
    \end{subfigure}
    \caption{{Impact of the loading condition upon the stress resolved avalanche statistics. (a) Probability density function (PDF) of stress-drop \(\Delta\tau\). (b) Complementary Cumulative Distribution Function of {stress-drops} \(\Delta\tau\) }}
    \label{fig:stressresolved_orientation}
\end{figure}

We also performed a stress resolved analysis of the avalanche statistics. Data are provided in Fig. \ref{fig:stressresolved_density} for [001] simulations at different initial dislocation density and Fig. \ref{fig:stressresolved_orientation} for various loading conditions. Avalanche statistics described by stress drops $\Delta \tau$ share many of the qualitative features of the avalanche statistics described by strain bursts $\Delta \gamma$. A well defined PWL regime can be seen over several orders of magnitudes of event sizes, and is delimited by cutoffs. To provide a quantitative et systematic sense of the evolution of avalanche statistics, all data were fitted using a truncated PWL equation with an exponential decay (see above). The PWL exponent is about -1.5 to -1.6 and is independent from dislocation density and loading orientation. The maximum cutoff decreases as the dislocation density increases as best seen on the CCDF curves.

\newpage

\bibliographystyle{unsrt}
\bibliography{biblio_formatted}
\end{document}